\documentclass[twocolumn,prc,amssymb, aps,superscriptaddress, showpacs,preprintnumbers,showkeys,
amsmath,floatfix]{revtex4}

\usepackage{color}
\usepackage{amsmath,mathtools,booktabs,
            microtype} %,siunitx}

\usepackage{epstopdf}
\usepackage{graphics}
\usepackage{graphicx}
\usepackage{dcolumn}
\usepackage{bm}
\usepackage{epsfig}
\usepackage{epstopdf}
\usepackage{times}
\usepackage{url}

\newcommand{\be}{\begin{equation}}
\newcommand{\ee}{  \end{equation}}
\newcommand{\ba}{\begin{eqnarray}}
\newcommand{\ea}{  \end{eqnarray}}
\newcommand{\ve}{\varepsilon}

\begin{document}

\title{Laser-Nucleus Interactions: The Quasiadiabatic Regime}

\author{Adriana \surname{P\'alffy}}
\email{palffy@mpi-hd.mpg.de}
\affiliation{Max-Planck-Institut f\"ur Kernphysik, Saupfercheckweg 1, D-69117 Heidelberg, Germany}

\author{Oliver \surname{Buss}}

\author{Axel \surname{Hoefer}}
\affiliation{AREVA GmbH, Dept.~Radiology $\&$ Criticality, Kaiserleistr. 29, D-63067 Offenbach am Main, Germany}

\author{Hans A. \surname{Weidenm\"uller}}
\email{haw@mpi-hd.mpg.de}
\affiliation{Max-Planck-Institut f\"ur Kernphysik, Saupfercheckweg 1, D-69117 Heidelberg, Germany}

\date{\today}

%%%%%%%%%%%%%%%%%%%%%%%%%%%%%%%%%%%%%%%%%%555
\begin{abstract}

The interaction between nuclei and a strong zeptosecond laser pulse
with coherent MeV photons is investigated theoretically. We provide a
first semi-quantitative study of the quasiadiabatic regime where the
photon absorption rate is comparable to the nuclear equilibration
rate. In that regime, multiple photon absorption leads
to the formation of a compound nucleus in the so-far unexplored regime
of excitation energies several hundred MeV above the yrast line. The
temporal dynamics of the process is investigated by means of a set of
master equations that account for dipole absorption, stimulated
 dipole emission, neutron decay and induced fission in a
chain of nuclei.  That set is solved numerically by means
of state-of-the-art matrix exponential methods also used in nuclear fuel burnup 
and radioactivity transport calculations. Our quantitative estimates predict the
excitation path and range of nuclei reached by neutron decay and
provide relevant information for the layout of future experiments.

\end{abstract}
%%%%%%%%%%%%%%%%%%%%%%%%%%%%%%%%%%%%%%%%%%%%%%

% 25.20.-x 	Photonuclear reactions
% 24.60.Dr 	Statistical compound-nucleus reactions
% 25.70.Gh 	Compound nucleus 
% 21.10.Pc 	Single-particle levels and strength functions 
% 42.50.Ct 	Quantum description of interaction of light and matter; related experiments
\maketitle

%%%%%%%%%%%%%%%%%%%%%%%%%%%%%%%%%55
\section{Introduction }
%%%%%%%%%%%%%%%%%%%%%%%%%%%%%%%%%%%%%%%%%%%%%

Recent experimental developments in laser physics promise to open the
new field of laser-induced nuclear reactions in a domain of excitation
energies that has not been explored so far. Efforts are under
way~\cite{Mou11} to generate a multi-MeV zeptosecond pulsed laser beam
at the Nuclear Physics Pillar of the Extreme Light Infrastructure
(ELI) now under construction in Romania~\cite{Eli12} and at the
International Center on Zetta-Exawatt Science and Technology
(IZEST)~\cite{Izest}. Furthermore, theoretical proposals for the
generation of coherent gamma-ray frequency combs at ELI have also been
put forward \cite{Kasia14}. How will an intense laser pulse interact
with a medium-weight or heavy target nucleus? The nucleus is a
strongly bound system. Therefore, the laser-nucleus interaction is
weak in comparison to the laser-atom one. A reaction that differs
significantly from the standard photon-induced nuclear reaction is
expected to occur only if the photons in the laser pulse are
coherent. Only then does the effective dipole width attain values in
the MeV range, making it comparable with other characteristic nuclear
energy scales. In this paper we accordingly consider the interaction
of a strong coherent zeptosecond laser pulse with a medium-weight or
heavy nucleus with mass number $A$. The pulse contains $N = 10^3 -
10^4$ coherent photons, the energy $E_L$ per photon is several MeV,
and the duration  of the pulse is $\hbar / \sigma$
where $\sigma$ is of the order of several $10$ keV so that $\hbar /
\sigma \approx 10^{- 20}$~s. Relevant questions then are: (i) How does
the interaction of this laser pulse differ from laser-matter
interaction in other areas of physics~\cite{Pia12}? (ii) Which are the
reactions that we expect to occur? The answers are obviously
interesting in their own right and relevant for the layout of future
experiments.

In the present paper we provide a partial answer to these questions by
addressing the quasiadiabatic regime of the laser-nucleus interaction.
In that regime, the process of photon absorption and that of nuclear
relaxation are governed by similar time scales. The paper follows the
study of the perturbative regime by one of the authors~\cite{Wei11}
where the absorption process is much slower than nuclear
relaxation. We hope to be able to address the sudden regime
(characterized by the converse situation) in a future paper. A brief
summary of first qualitative results for the quasiadiabatic regime has
been published in Ref.~\cite{Pal14}.

Our approach is based on the master equation describing the excitation
and relaxation of the nucleus under the influence of the external
field provided by the laser. Multiple absorption of coherent photons
leads to nuclear excitation far above yrast. Setting up the master
equation requires, therefore, the knowledge of the $A$-particle level
density $\rho_A$ at high excitation energies and for large particle
numbers $A$, expressed in terms of the single-particle level density
$\rho_1$. That is a challenging problem because near its maximum and
for $A \gtrsim 100$, $\rho_A$ is several tens of orders of magnitude
larger than $\rho_1$.  An important preparatory step in our work has
been the construction of a reliable approximation for $\rho_A$ in
terms of $\rho_1$~\cite{Pal13a, Pal13b}.

Use of the master equation renders possible the semi-quantitative study
of the competition between photon absorption, stimulated photon
emission, photon-induced nucleon emission, neutron evaporation, and
induced fission. In the absence of particle emission and fission,
photon absorption would saturate at an excitation energy where the
widths for absorption and for stimulated emission become equal. That
is the case at the energy where the level density $\rho_A$ reaches its
maximum. Neutron evaporation takes over at an energy below the
saturation point. The combination of repeated neutron
  emission and continued dipole absorption by the daughter nuclei then
  produces proton-rich nuclei far from the valley of stability.
Although the induced fission width is small in comparison to all other
widths, fission eventually terminates the reaction chain unless the
laser pulse comes to an end beforehand. In the latter case,
laser-nucleus interaction experiments promise to shed light on nuclei
at excitation energies far above yrast and far from stability.

A qualitative description of the expected processes and a definition
of the quasiadiabatic regime are given in Sec.~\ref{gen}. The master
equation and the transition rates are introduced in Sec.~\ref{mas}.
This section also contains a semi-quantitative estimate of the energies
and time scales involved in the photon-absorption, neutron-evaporation
and fission processes. Numerical results follow in Sec.~\ref{numres}
and the paper concludes  with a discussion in
Sec.~\ref{disc}.

%%%%%%%%%%%%%%%%%%%%%%%%%%%%%%%%%%%%%%%%%%%%%
\section{General Considerations}
\label{gen}
%%%%%%%%%%%%%%%%%%%%%%%%%%%%%%%%%%%%%%%%%%%%%

Nuclei are bound by the strong interaction. As a consequence, even the
interaction of the strong laser pulse defined in the Introduction with
a nucleus is much less violent than the interaction of a
medium-intensity optical laser pulse with an atom. We substantiate
that statement with the help of the Keldysh parameter~\cite{Kel65}
$\gamma = \sqrt{2mI} \omega / (e \mathcal{E})$ used in atomic
physics. Here, $m$ and $e$ are mass and charge of the electron,
respectively, $I$ is the field-free ionization potential, and $\omega$
and $\mathcal{E}$ are the frequency and electric field strength,
respectively, of the laser pulse. The Keldysh parameter determines the
dominant interaction mechanism in atoms. For $\gamma < 1$ tunneling
ionization dominates while for $\gamma \gg 1$ the process is governed
by multiphoton ionization. A small value of $\gamma$ corresponds in
the optical regime to an electric field strength $\mathcal{E} \approx
10^9$ eV/cm of the laser. Such a field distorts the Coulomb potential
of the atom so strongly that electrons are set free. The nuclear
equivalent of the Keldysh parameter is obtained by replacing $m$ by
the nucleon mass ($m \to 2000 \ m$), $\omega$ by the photon energy
($\omega \to 10^6 \ \omega$), and $I$ by the binding energy of the
last nucleon ($I \to 10^7 \ I$). These substitutions increase the
value of $\gamma$ by a factor $10^{11}$. To return $\gamma$ to a value
less than unity, the field strength would have to increase by that
same factor $10^{11}$, i.e., $\mathcal{E}$ would have to be of the
order $10^{20}$ eV/cm. This value roughly corresponds to the ratio of
the binding energy of the last nucleon and the nuclear radius and,
thus, to a distortion of the nuclear potential that roughly
corresponds to the above-mentioned distortion of the Coulomb potential
in atoms. In comparison with a standard laser, the photon energy in
our laser pulse is increased by six orders of magnitude. Such an
increase falls short by a wide margin of the necessary increase of
$\mathcal{E}$ by eleven orders of magnitude. Therefore, the
laser-nucleus interaction is governed by a value $\gamma \gg 1$ of the
Keldysh parameter. That confirms our initial statement that even a strong
laser pulse perturbs the nucleus only fairly weakly. In what follows
we actually focus on sequential absorption of single photons within one pulse, i.e.,
on even weaker electric field strengths than required for
simultaneous multiphoton absorption. In
  contrast to atomic physics, in nuclei  an additional time
  scale plays an important role, namely the nuclear relaxation time (see below). In the
  quasiadiabatic regime that time scale is comparable with the time
  scale for absorption of a single photon. Single-photon absorption
  and nuclear relaxation occur therefore more or less simultaneously.

For photons in the MeV range, the product of photon wave number $k$
and nuclear radius $R$ obeys $k R \ll 1$. Therefore, we consider only
dipole processes even though quadrupole excitation is important for
some nuclei at small excitation energies~\cite{Pal08}. For $N \gg 1$
coherent photons in the laser pulse, dipole excitation is governed by
the effective dipole width $N \Gamma_{\rm dip}$. Here $\Gamma_{\rm
  dip}$ is the standard nuclear dipole width and is in the keV
range. The amplification factor $N$ applies in the semiclassical limit:
 For $N$ coherent photons the matrix element for
  absorption of a single photon is proportional to $\sqrt{N}$, and the
  rate is proportional to $N$. After absorption of $N_0$ photons the rate
  becomes proportional to $(N - N_0)$. For $N \gg N_0$ we may simply use $N-N_0\approx N$. For $N
\Gamma_{\rm dip}$ we use values around $5$ MeV.  Coherence is vital in
bringing the effective dipole width up to values that are comparable
with other characteristic nuclear energy scales defined below. Without
coherence, the probability for the processes investigated in this
paper would be dramatically reduced. In the course of the reaction, up
to $N_0 \approx 5 \times 10^2$ photons may be absorbed. 

A further distinguishing feature of nuclei is that the nucleon-nucleon
force is basically attractive, and that nuclei are self-bound. Nuclear
properties are understood in terms of the shell-model potential (the
mean field) plus the remaining ``residual'' nucleon-nucleon
interaction. Because of the latter, distinct modes of nuclear
excitation have the tendency to mix with the numerous other nuclear
modes that are near the same excitation energy: the nucleus
equilibrates. That property is absent in atoms. It qualitatively
changes the treatment of the multistep photon absorption process. The
time scale $\hbar / \Gamma_{\rm sp}$ for equilibration is expressed
in terms of the spreading width $\Gamma_{\rm sp}$, a manifestation of
the residual interaction. The value of $\Gamma_{\rm sp}$ is known for
low-lying modes with excitation energies of up to ten or twenty MeV
where $\Gamma_{\rm sp}$ is of order $\Gamma_{\rm sp} \approx 5$
MeV~\cite{Wei09}.

The ratio $N \Gamma_{\rm dip} / \Gamma_{\rm sp}$ relates the speed of
dipole absorption with that of nuclear relaxation and, therefore,
defines three regimes of the laser-nucleus interaction. (i) In the
perturbative regime $N \Gamma_{\rm dip} \ll \Gamma_{\rm sp}$, single
excitation of the collective dipole resonance in nuclei dominates.
That regime has been investigated in Ref.~\cite{Wei11}. Consequences
for future experiments were theoretically explored in
Ref.~\cite{Die10}. It was shown that if excited above neutron
threshold, the time dependence of the nuclear decay is
non-exponential, both in the neutron and in the gamma decay
channels. (ii) In the sudden regime $N \Gamma_{\rm dip} \gg
\Gamma_{\rm sp}$ the residual nucleon-nucleon interaction is
irrelevant. Nucleons are excited independently of each other and are
emitted from the common average shell-model potential. The potential
readjusts after nucleon emission. If the duration time $\hbar /
\sigma$ of the laser pulse is sufficiently large, the nucleus
evaporates. The sudden regime is so far unexplored. (iii) The
quasiadiabatic regime $N \Gamma_{\rm dip} \approx \Gamma_{\rm sp}$
forms the topic of this paper. The nucleus (almost) attains
statistical equilibrium between any two subsequent photon absorption
processes.

A quasiadiabatic process occurs when the energy $E_L$ per photon is
less than or comparable with the nucleon binding energy $E_B$ of
around $8$ MeV. Then absorption of a single photon does not lead to
nucleon emission. Rather, the excitation energy is shared (almost)
instantaneously with several or many other nucleons. The nucleus
equilibrates. Another photon is absorbed, and the process repeats
itself. Consecutive absorption of $N_0 \gg 1$ photons leads to high
excitation energies $N_0 E_L$ of the (almost) equilibrated compound
nucleus. The excitation process terminates either with the laser pulse
or when all nuclei have fissioned. For the duration time $\hbar /
\sigma$ of the laser pulse, significant emission of particles sets in
only after several or even many photons have been absorbed.
Precompound reactions show that the condition $E_L \leq E_B$ is not
absolutely necessary \cite{Mit10}. If equilibration is sufficiently
fast, a quasiadiabatic process will occur also for $E_L > E_B$. 

With  the nucleus close to
equilibrium at all times, photon absorption leads to
compound-nucleus formation at very high energies. That situation is
distinctly different from Coulomb excitation~\cite{Ber08} or nuclear
excitation by inelastic electron scattering~\cite{Ben08}. In both
these processes, modes of excitation very far from the equilibrated
compound nucleus are formed at some excitation energy $E$. Such
modes decay by particle emission and/or equilibration. If reached at
all, the compound nucleus is formed at a much lower excitation
energy than $E$.

The equilibration mechanism being absent in atoms, our theoretical
description differs from the strong-field approximation in atomic
physics~\cite{Mil06,Popr14} and is related to the theory of
precompound reactions~\cite{Wei08}. As mentioned in the Introduction,
we describe the process in terms of a set of time-dependent master
equations. In addition to dipole absorption, we take stimulated dipole
emission, neutron evaporation,  and fission into account.
Induced particle emission is briefly addressed in Section\ref{disc}.

We simplify the treatment in two respects. (i) We disregard spin
altogether. (ii) We assume that after each photon absorption process,
the nucleus attains full equilibrium. Both approximations simplify our
treatment considerably. Without approximation (i), the number of
master equations would be multiplied by the number of spin values
considered. Without approximation (ii), the same would happen with
regard to the number of configurations needed to describe equilibration
at fixed excitation energy. While both approximations can easily be
removed, the resulting increased complexity of the approach does not
seem justified at this early stage of investigation and in view of
the complete lack of experimental data.

\begin{figure}[ht]
\includegraphics[width=0.9\linewidth]{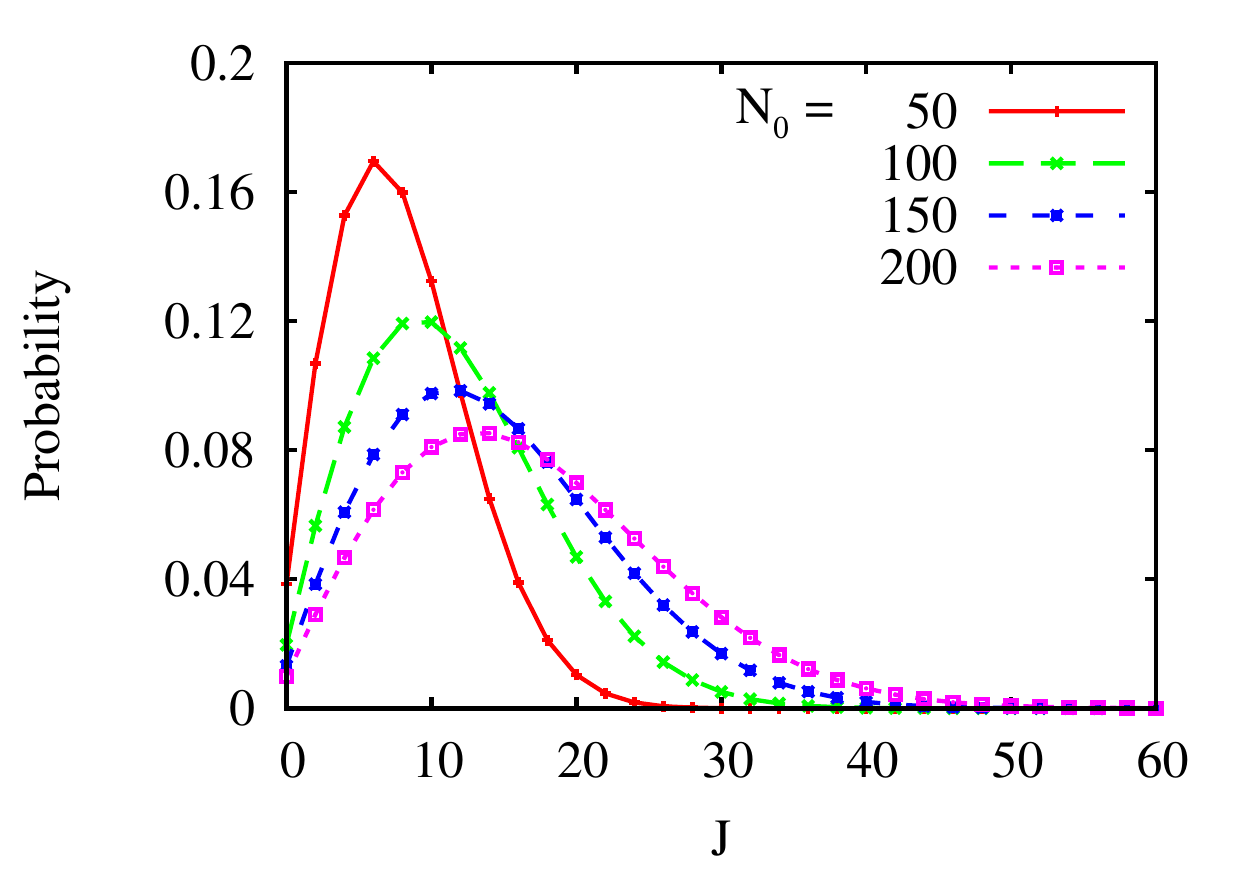}
\caption{(color online).  Distribution of total spin
    values of the compound nucleus for several values of the number
    $N_0$ of absorbed photons.}
\label{fig1}
\end{figure}

To justify approximation (i), we calculate analytically in
Appendix~\ref{A} the distribution of nuclear spin values after $N_0$
photons have been absorbed by dipole transitions by a nucleus with
ground state spin zero. This is done for $N_0 \gg 1$. The distribution
peaks at spin $J = \sqrt{N_0}$ and falls off very rapidly for larger
values of $J$. Exact numerical results not based on the approximation $N_0
\gg 1$ confirm this result, see Fig.~\ref{fig1}. Even for a maximum
number $N_0 = 5 \times 10^2$ of absorbed photons, the nuclear spin
does not significantly exceed $20$ or so. This fact justifies our
neglect of spin.

Approximation (ii) implies that in the master equation we need not
distinguish (as done in precompound reactions) the formation of
$n$-particle $n$-hole states. Instead we work with the full set of
equilibrated states at each excitation energy. To discuss the error
made in approximation (ii), we observe that for a non--equilibrated
system the mean number of $n$--particle $n$--hole pairs at fixed
excitation energy is smaller than for the equilibrated system. That
has two consequences. First, the number of states accessible for
further dipole excitation is larger (because the exclusion principle
blocks fewer states). Second, the mean excitation energy per particle
or hole is larger, too. Therefore, neutron decay is more likely than
in the equilibrated case (the number of available decay channels is
increased). Both errors work in the same direction, leading within our
approximation to an overestimate of the overall duration time of the
excitation process. The resulting uncertainty can be compensated by
varying the relative strength of photon excitation and neutron decay
in the calculations.

%%%%%%%%%%%%%%%%%%%%%%%%%%%%%%%%%%%%%%%%%%%%%%%%%%%%%%%%%%%%%%%%%%%%%
\section{Master Equation}
\label{mas}
%%%%%%%%%%%%%%%%%%%%%%%%%%%%%%%%%%%%%%%%%%%%%%%%%%%%%%%%%%%%%%%%%%%%%

%%%%%%%%%%%%%%%%%%%%%%%%%%%%%%%%%%%%%%%%%%%%%%%%%%%%%%%%%%%%%%%%%%%%%
\subsection{Basic Approach}
%%%%%%%%%%%%%%%%%%%%%%%%%%%%%%%%%%%%%%%%%%%%%%%%%%%%%%%%%%%%%%%%%%%%%

With $A$ the mass number of the target nucleus, we consider a chain of
$(n + 1)$ nuclei with mass numbers $A - i$ where $i = 0, 1, 2, \ldots,
n$, with an arbitrary cutoff at $i = n$.  In nuclei with
  even mass number $A$, the states with spin zero at excitation energy
  between $(k - 1/2) E_L$ and $(k + 1/2) E_L$ are grouped together and
  are jointly referred to as states $(i, k)$. Here $E_L$ is the photon
  energy and $k = 1, 2, \ldots$. The group of states with excitation
  energies in the interval $0 \leq E \leq (1/2) E_L$ is labeled $(i,
  0)$. The number of such states is determined by the level density
  $\rho(A, E)$ of states with spin zero. For odd $A$ we proceed
  analogously. For simplicity and in order to avoid the introduction
  of additional parameters we neglect the even-odd staggering of the
  ground-state energies as well as the spin-cutoff factor, and
  approximate the level density of spin $1/2$ states by interpolating
  between the values for the two neighboring even $A$ nuclei. In other
  words, we use the expression for $\rho(A, E)$ valid for even $A$ and
given in Ref.~\cite{Pal13b} indiscriminately for both even and odd
$A$. We construct the time-dependent master equation for the total
  occupation probability $P(i, k, t)$ of these states as function of
  time $t$. The equation takes into account dipole excitation by the
  coherent laser pulse, stimulated dipole emission by the same pulse,
  both for every nucleus in the chain, and neutron decay populating
  nucleus $A - i - 1$ at the expense of nucleus $A - i$. It is assumed

  that within each group $k$ of states in nucleus $A - i$, the
  occupation probability is equilibrated at all times and, thus,
  proportional to the total level density $\rho(i, k)$ for that group.
  That assumption is characteristic of the quasiadiabatic
  regime. Neglecting the emission of charged particles we confine
  ourselves to a chain of nuclei with equal proton numbers. Likewise
  we do not take account of particle loss due to direct photon
  excitation of particles into continuum states. We address the
  ensuing limitations and possible corrections below.

For the duration time $1 / \sigma$ of the laser pulse, the states $(i,
k)$ are fed by coherent dipole excitation of the states $(i, k - 1)$
and by stimulated dipole emission of the states $(i, k + 1)$, and they
are depleted by dipole absorption exciting the states $(i, k + 1)$ and
by stimulated dipole emission to the states $(i, k - 1)$. Using
Fermi's golden rule we write the rates feeding the states $(i, k)$ as
$W_{k' k}^2 \rho(i, k)$ with $k' = k \pm 1$. Here $W^2_{k k'} =
W^2_{k' k}$ is the square of the transition matrix
element. Neutron decay depletes the states $(i, k)$ at the rate
$\Gamma_N(i, k)$. Neutron decay of the states $(i - 1, k')$ in the
nucleus with mass number $A + 1 - i$ feeds the states $(i, k)$ with
the rate $\Gamma_N(i - 1, k' \to k)$. The master equation reads
\begin{widetext}
\ba
\dot{P}(i, k, t) &=& \Theta(1/\sigma - t) \bigg\{ \rho(i, k) [
  W^2_{k k - 1} P(i, k-1, t) + W^2_{k k + 1} P(i, k + 1, t) ] - P(i,
k, t) [ W^2_{k k - 1}\rho(i, k - 1) + \nonumber \\ &+& W^2_{k k + 1}
  \rho(i, k + 1) ] \bigg\} + \sum_{k'} \Gamma_N(i - 1, k' \to k) P(i -
1, k', t) - \Gamma_N(i, k) P(i, k, t) \ .
\label{1}
\ea
\end{widetext}
Here, $\hbar = 1$ and we have defined $P(- 1, k, t) = 0$. The dot
denotes the time derivative, and $\Theta$ is the Heaviside function.
The initial condition is $P(i, k, 0) = \delta_{i 0} \delta_{k 0}$. We
require that neutron emission does not take place from the nucleus
with mass number $A - n$ and put $\Gamma_N(n, k) = 0$ for all $k$ so
that nucleus $(A - n)$ serves as a dump for the overall probability
flow. For $i = 0, 1, \ldots, n - 1$ we have
\be
\Gamma_N(i, k) = \sum_{k'} \Gamma_N(i, k \to k') \ .
\label{2}
\ee
Then Eq.~(\ref{1}) implies $\sum_{i, k} \dot{P}(i, k, t) = 0$, and the
master equation conserves total occupation probability.

Induced fission is taken into account by introducing a diagonal loss
term $-\Gamma_f(i,k) P(i, k, t)$ in Eq.~(\ref{1}). Here
$\Gamma_f(i,k)$ is the width for induced fission from state $(i, k)$.
Fission leads to a depletion of the total occupation probability. We
do not keep track of the fission products. Therefore, fission
eventually terminates the laser-nucleus reaction. In the absence of
induced fission and neutron decay (i.e., for $\Gamma_N(i, k) =
\Gamma_f(i,k) = 0$ for all $(i, k)$), probability is conserved within
the target nucleus, $\sum_k \dot{P}(0, k, t) = 0$. For $\sigma \to 0$
(infinitely sharp laser energy with the laser pulse lasting forever)
and $\Gamma_N(i, k) = 0$ for all $(i, k)$, the target nucleus
equilibrates, asymptotically ($t \to \infty$) reaching the equilibrium
distribution $P_{\rm eq}(0, k) \propto \rho(i, k)$ for all $k$ values
below and around the saturation energy.

%%%%%%%%%%%%%%%%%%%%%%%%%%%%%%%%%%%%%%%%%%%%%%%%%%%%%%%%%%%%%%%%%%%%%&
\subsection{Transition Rates}
\label{rates}
%%%%%%%%%%%%%%%%%%%%%%%%%%%%%%%%%%%%%%%%%%%%%%%%%%%%%%%%%%%%%%%%%%%%%%

The transition rates have been defined, calculated, and discussed in
Ref.~\cite{Pal14}. We present these rates here for the sake of
completeness. Because of the approximations explained below in the
calculation of the level density, particular values of the transition
rates might differ from the ones presented in Ref.~\cite{Pal14}. Since
$\hbar = 1$ we use the expressions ``width'' and ``rate''
interchangeably.

%%%%%%%%%%%%%%%%%%%%%%%%%%%%%%%%%%%%%%%%%%%%%%%%%%%%%%%%%%%%%%%%%%%%%
\subsubsection{Dipole Transitions}
%%%%%%%%%%%%%%%%%%%%%%%%%%%%%%%%%%%%%%%%%%%%%%%%%%%%%%%%%%%%%%%%%%%%%

With $\Gamma_{\rm dip}$ the standard nuclear dipole absorption width
and $N$ the number of coherent photons in the laser pulse, the
effective dipole width for the ground state is given by $N \Gamma_{\rm
  dip}$. That expression holds for $N \gg 1$ in the semiclassical
approximation. With $\Gamma_{\rm dip}$ in the keV range and $N \approx
10^3$ or $10^4$, the effective dipole width is of the order of several
MeV. The value of $N \Gamma_{\rm dip}$ serves as an input parameter
for our calculation. Photon absorption of an equilibrated compound
nucleus at excitation energy $E$ is then governed by the effective
absorption rate $(N \Gamma)_{\rm eff}(E) = N \Gamma_{\rm dip}
\ \rho_{\rm acc}(E) / \rho_{\rm acc}(E_g)$. Here $\rho_{\rm acc}(E)$
is the density of accessible states and $E_g$ is the energy of the
ground state. The expression for $(N \Gamma)_{\rm eff}(E)$ is valid as
long as the number $N_0$ of absorbed photons is small compared to
$N$. That is the case for the calculations presented below. We equate
$(N \Gamma)_{\rm eff}(k E_L)$ (the rate for population of the states
$(i, k)$ by dipole absorption of the states $(i, k - 1)$) with $W^2_{k
  (k - 1)} \rho(i, k)$ in Eq.~(\ref{1}). The stimulated dipole
emission width for the inverse transition $(i, k) \to (i, k - 1)$ is
then given by detailed balance as $(N \Gamma)_{\rm st}(k E_L) = N
\Gamma_{\rm eff}(k E_L) \ \rho(i, k - 1) / \rho(i, k)$. In this way,
all dipole rates in Eq.~(\ref{1}) are determined.  With $\rho(i, E)$
the level density of nucleus $(A - i)$ at excitation energy $E$, we
approximate the density of states $(i, k)$ as
$\rho(i,k)=\rho(i,kE_L)$.  

The level densities $\rho(A,E)$ are calculated using the expressions
for the total level density of spin-zero states in nucleus $A$ as a
function of excitation energy $E$ derived in Ref.~\cite{Pal13b} as
functions of the single-particle level density $\rho_1(\ve)$. A
realistic linear or quadratic energy dependence of $\rho_1$ is
considered,
\be
\rho^{(1)}_1(\ve) = \frac{2 A}{F^2} \ve \ , \ \rho^{(2)}_1(\ve) =
\frac{3 A}{F^3} \ve^2 \ .
\label{4}
\ee
Here $V$ with $0 \leq \ve \leq V$ defines the range of the
single-particle spectrum, and $F$ is the Fermi energy. The
single-particle energies $\ve_i$ with $i = 1, 2, \ldots$ are obtained
from the condition $i = \int_0^{\ve_i} {\rm d} \ve' \ \rho_1(\ve')$.
The expressions~(\ref{4}) are approximately valid for $A = 100$ and $A
= 200$, respectively. We have chosen $V = 45$ MeV and $F = 37$
MeV  and keep these values constant throughout the 
neutron decay chain. These values also determine the chosen number of total
bound single-particle states \cite{Pal13b}, namely 148 for $A=100$ and 360 for $A=200$, 
respectively. 

When the number of nucleons is large, the method~\cite{Pal13b} fails
to work at small excitation energies. In this region we use the
Bethe formula~\cite{Bet36}. This is done for the first 65 MeV of the
excitation energy for the case of medium-weight nuclei ($A = 100$) and
the first 200 MeV for heavy nuclei ($A=200$), i.e., for approximately
10$\%$ of the total relevant spectrum. For the density $\rho_{\rm
  acc}(E)$ of accessible states we have used the Fermi-gas model
described in Ref.~\cite{Pal13b} and the same choices of $\rho_1$ as in
Eq.~(\ref{4}).

Figures ~\ref{fig2} and ~\ref{fig3} give the widths for effective
dipole absorption, stimulated dipole emission, neutron emission and
the induced fission rates for medium-weight ($A=100$) and heavy
($A=200$) nuclei, respectively. Calculation of the latter two rates is
explained below. Figures~\ref{fig2} and \ref{fig3} show that for both
medium-weight ($A=100$) and heavy nuclei ($A=200$), the effective
dipole absorption width $(N \Gamma)_{\rm eff}(E)$ decreases slowly
with increasing excitation energy $E$. Over a range of $1000$ MeV the
decrease amounts typically to a factor of two. The stimulated dipole
emission width $(N \Gamma)_{\rm st}(E)$, also shown in
Figs.~\ref{fig2} and \ref{fig3}, starts out at small excitation energy
from a value much below that of $(N \Gamma)_{\rm eff}(E)$ and
increases monotonically with $E$. It becomes equal to $(N \Gamma)_{\rm
  eff}(E)$ at the maximum $E_{\rm max}$ of the level density $\rho(k)$. 
Significant dipole excitation above the energy $E_{\rm max}$ is not
possible because stimulated emission outweighs here absorption, $(N
\Gamma)_{\rm st}(E) > (N \Gamma)_{\rm eff}(E)$ for $E > E_{\rm
  max}$. For the two choices of $\rho_1$ in Eq.~(\ref{4}) we have
$E_{\rm max} = 533$ MeV and $E_{\rm max} = 1200$ MeV, respectively.

We have tacitly assumed that the spreading width $\Gamma_{\rm sp}$
retains the value of $\approx 5$ MeV found at low excitation energy
also at the much higher excitation energies relevant for our paper. As
is the case for $(N \Gamma)_{\rm eff}(E)$, the energy dependence of
the spreading width is determined by the relevant density of
accessible states. As shown for $(N \Gamma)_{\rm eff}(E)$ in
Figs.~\ref{fig2} and \ref{fig3}, that density changes only moderately
with excitation energy. That fact validates our assumption.

%%%%%%%%%%%%%%%%%%%%%%%%%%%%%%%%%%%%%%%%%%%%%%%%%%%%%%%%%%%%%%%%%%%%%%
\subsubsection{Neutron Decay}
%%%%%%%%%%%%%%%%%%%%%%%%%%%%%%%%%%%%%%%%%%%%%%%%%%%%%%%%%%%%%%%%%%%%%%

For the neutron decay rates we use the Weisskopf estimate
\be
\Gamma_N(i, k) = \frac{1}{2 \pi \rho(i, k)} \int_0^{(k + 1/2) E_L
- B_N(i)} {\rm d} E' \ \rho(i + 1, E') \ , \nonumber \\
\ee
\be
\Gamma_N(i, k \to k') = \frac{1}{2 \pi \rho(i, k)} \int_{(k' -
1/2) E_L}^{(k' + 1/2) E_L} {\rm d} E' \ \rho(i + 1, E') \ .
\label{5}
\ee
Here $B_N(i)$ is the neutron binding energy of nucleus $(A - i)$. In
the second of Eqs.~(\ref{5}) the lower bound is zero for $k' = 0$, and
$k'$ is bounded by $k' E_L \leq k E_L - B_N(i)$. Eqs.~(\ref{5}) are
consistent with Eq.~(\ref{2}). We simplify the calculations by
assuming that $B_N(i) = B_N = 8$ MeV for all $i = 0, 1, 2, \ldots, n$
for the short nuclear chains considered. We thereby neglect the
odd-even staggering of binding energies and level densities. These run
in parallel and, therefore, largely compensate each other in the
neutron decay widths. For both choices for the single-particle level
density $\rho_1$ in Eq.~(\ref{4}), Figs.~\ref{fig2} and \ref{fig3}
show that $\Gamma_N(k)$ rises steeply with excitation energy. While
much smaller than $(N \Gamma)_{\rm eff}$ for small excitation
energies, $\Gamma_N(k)$ becomes equal to $(N \Gamma)_{\rm eff}$ at $E
= E_N$ and exceeds $(N \Gamma)_{\rm eff}$ for $E > E_N$. For the two
choices of $\rho_1$ and $T(E) = 1$ we have $E_N \approx 435$ MeV and
$E_N \approx 1080$ MeV, respectively. Both values are smaller than the
corresponding values of $E_{\rm max}$, $E_{\rm max} = 533$ MeV and
$E_{\rm max} = 1200$ MeV.  We note that the crossing energies $E_N$
  are here larger than the ones presented in Ref.~\cite{Pal14}, i.e.,
  the neutron rates grow more slowly with excitation energy in the
  present case. That difference results from our treatment of the
  Fermi energy $F$. In Ref.~\cite{Pal14}, the Fermi energy changes
  with mass number $A$ whereas in the present work, we consider $F$
  fixed. That is in accordance with the fixed value of $B_N(i) = B_N =
  V - F =8$ MeV. The difference $E_{\rm max} - E_N$ is sufficiently
large in both cases and for both $A=100$ and $A=200$ to be physically
significant, in spite of the inherent uncertainties of the calculation
of level densities at high excitation energies. We conclude that
neutron evaporation is the limiting factor in nuclear excitation by
dipole absorption (provided that fission and proton decay can be
neglected).

In Eqs.~(\ref{5}) we have not taken account of the transmission
coefficients $T(E)$, i.e., the probability of formation of a residual
nucleus at excitation energy $E$ under neutron emission. These obey $0
\leq T(E) \leq 1$ and should multiply the integrands in
Eqs.~(\ref{5}).  Neutron decay is dominated by $s$-wave neutrons. Here
and except for the slowest neutrons the transmission coefficients are
of order unity~\cite{Aue62}. We test the influence of
that approximation (which overestimates the neutron widths) by
multiplying all neutron widths by a common factor  $1/2$. The pairs of
dark blue dashed lines in Figs.~\ref{fig2} and \ref{fig3} show the
band of neutron widths for $T(E)$ in the interval $1 / 2 \leq T(E)
\leq 1$. For $T(E) = 1/2$, $E_N$ is
  closer to $E_{\rm max}$.

%%%%%%%%%%%%%%%%%%%%%%%%%%%%%%%%%%%%%%%%%%%%%%%%%%%%%%%%%%%%%%%%%%%%
\subsubsection{Fission}
%%%%%%%%%%%%%%%%%%%%%%%%%%%%%%%%%%%%%%%%%%%%%%%%%%%%%%%%%%%%%%%%%%%

\begin{figure}[ht]
\includegraphics[width=\linewidth]{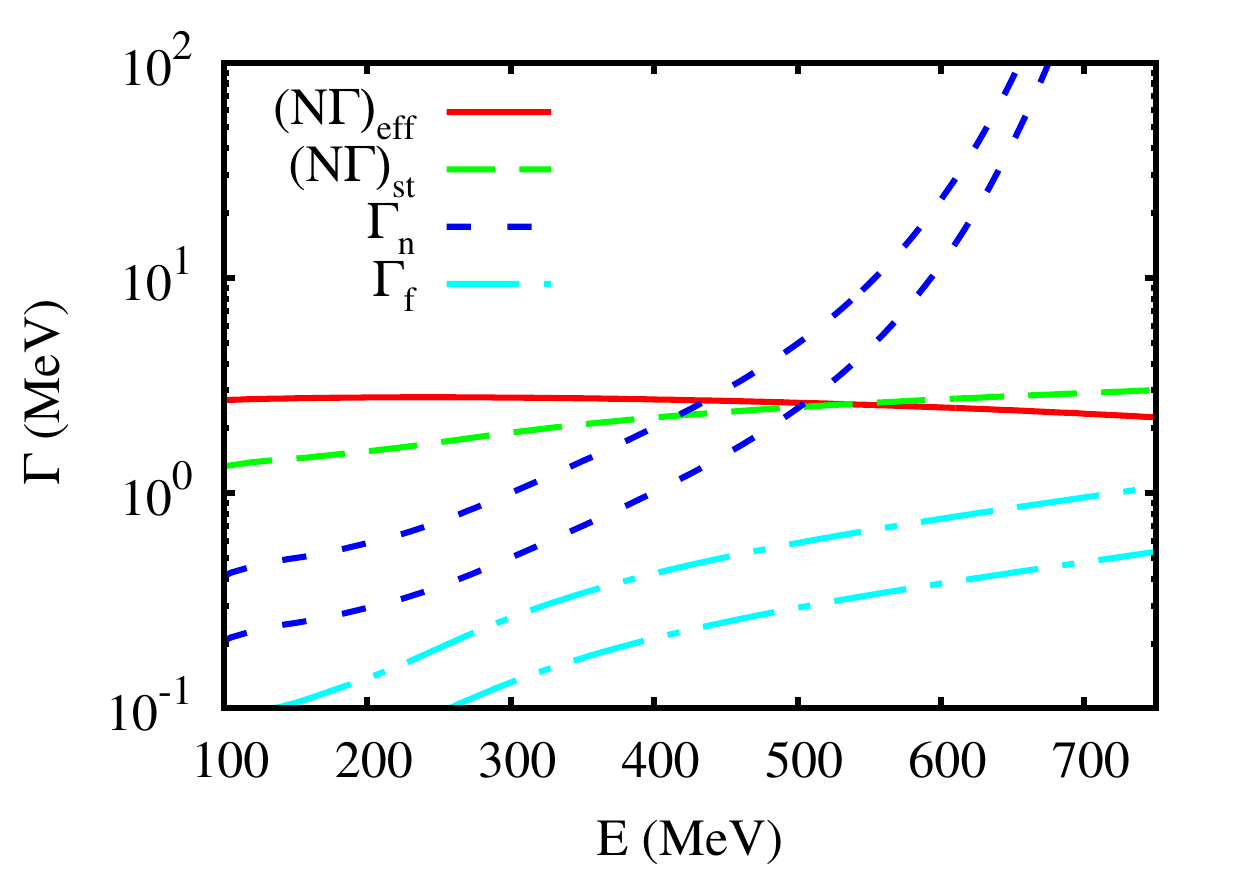}
\caption{(color online). Widths for effective dipole absorption (solid
  red line), stimulated dipole emission (long-dashed green line) (both
  for $N\Gamma_{\rm dip} = 5$ MeV), band for neutron emission
  (short-dashed dark blue lines) and band for induced fission
  (dashed-dotted light blue lines) versus excitation energy $E$ for $A
  = 100$ (148 bound single-particle states). The two neutron decay
  widths are calculated from the Weisskopf estimate ($T(E) = 1$) and
  by taking for the transmission coefficients uniformly the value
  $T(E) = 1/2$. The two fission widths are obtained using the values
  $\hbar\omega_1 = 2$ and 4 MeV. }
\label{fig2}
\end{figure}

\begin{figure}[ht]
\includegraphics[width=\linewidth]{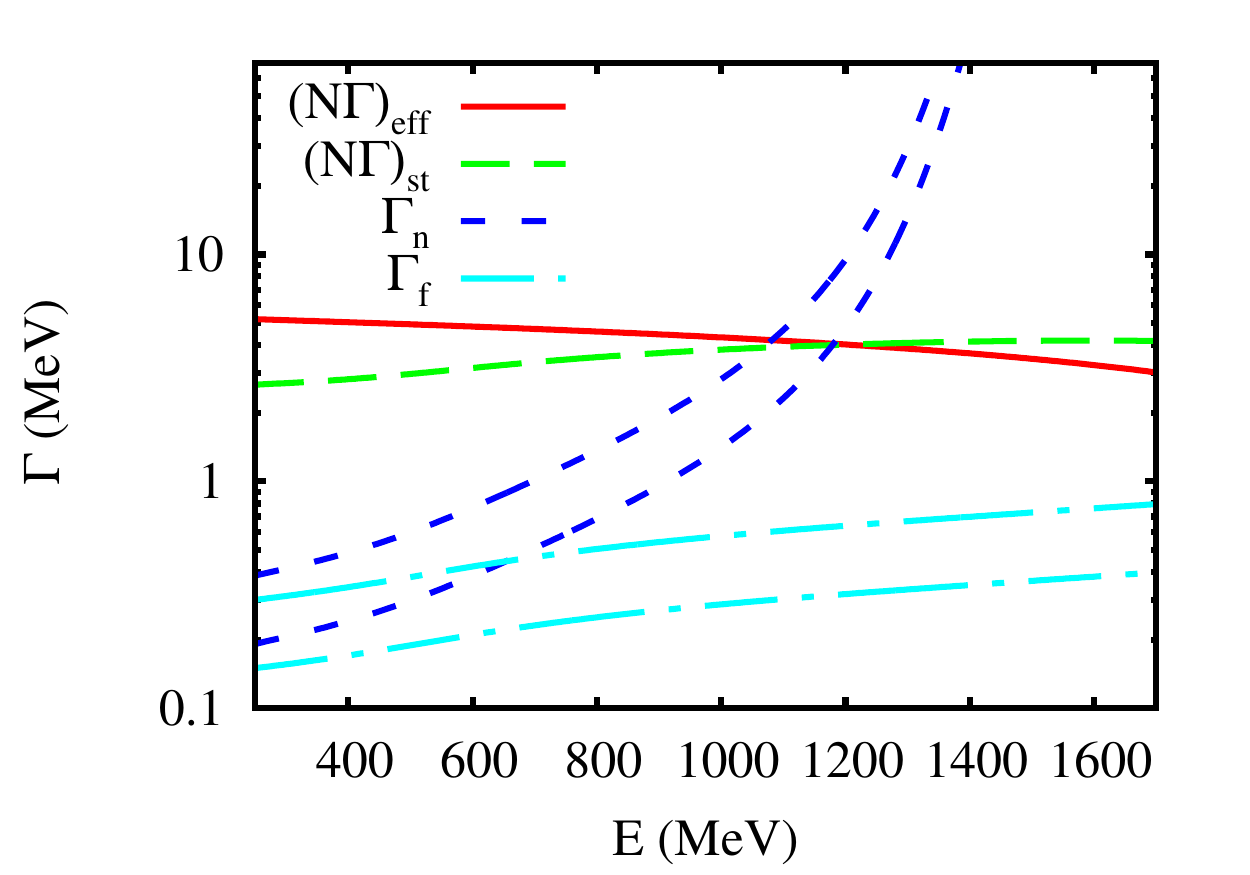}
\caption{(color online).  Same widths as in
 Fig.~\ref{fig2} for $A = 200$ (360 bound single-particle states).}
\label{fig3}
\end{figure}

According to the Bohr-Wheeler formula~\cite{Boh39} modified by
friction~\cite{Gra83}, the maximum width for induced fission (reached
at friction constant $\beta = 0$) depends on excitation energy $E$
essentially as
\be
\Gamma_f(i,E) = (\hbar \omega_1 / (2 \pi))\exp \{ - E_f / T \} \ .
\label{fiss}
\ee
Here, $E_f$ is the height of the fission barrier, $\omega_1$ is the
frequency of the inverted harmonic oscillator that osculates the
fission barrier at its maximum, and $T$ is the nuclear temperature.
With $T^{- 1} =({\rm d} / {\rm d} E) \ln \rho(i,E)$, the fission width
increases very slowly with $E$. That is shown in Figs.~\ref{fig2}
and \ref{fig3} where we present the calculated fission rates for
$\hbar\omega_1 = 4$~MeV and 2 MeV and for $E_f = 10$~MeV (4 MeV),
respectively. The two values of $E_f$ correspond to $A = 100$ ($A =
200$), respectively, the fission barrier being lower for heavy nuclei.
We see that while perhaps competitive with $(N \Gamma)_{\rm eff}$ at
low excitation energy and for very heavy nuclei, the fission width at
higher excitation energy is never competitive with dipole absorption
or neutron decay, see Ref.~\cite{Pal14}. We must keep in mind,
however, that for a sufficiently long laser pulse the processes
described by the master equation~(\ref{1}) are ultimately terminated
by fission. The induced fission width for $E = 200$ MeV for a
medium-weight nucleus, for instance, corresponds to a half-life of
approximately~5~zs. As shown in the following section by our numerical
results, fission is therefore terminating the laser-nucleus reaction
after several tens of zs.

Actually, little is known about the dependence of $\omega_1$ and $E_f$
on $E$ for excitation energies in the range of several $100$ MeV above
yrast. We expect $E_f$ to decrease with temperature and, more
significantly, as ever more neutrons are evaporated. That will speed
up the fission process. We have not attempted to estimate the
dependence of $E_f$ on temperature and mass number. Such an estimate
would require different techniques and is beyond the scope of this
paper.


%%%%%%%%%%%%%%%%%%%%%%%%%%%%%%%%%%%%%%%%%%%%%%%%%%%%%%%%%%%%%%%%%%%%
\subsection{Estimates \label{estimates}}
%%%%%%%%%%%%%%%%%%%%%%%%%%%%%%%%%%%%%%%%%%%%%%%%%%%%%%%%%%%%%%%%%%%%

Before presenting our numerical results we give some estimates that
show roughly what to expect. We estimate the time dependence of the
mean excitation energies and the range of nuclei reached by the
combination of multiple photon absorption and neutron decay. We recall
that the total neutron decay width $\Gamma_N(E)$ increases steeply
with excitation energy $E$ while the effective dipole absorption width
$(N \Gamma)_{\rm eff}(E_N)$ decreases slowly with $E$. The point $E_N$
of intersection of the two curves depends on $(N \Gamma)_{\rm eff}(E)$
only extremely weakly. For our estimate we therefore use the values
$E_N \approx 435$ MeV and $E_N \approx 1080$ MeV given in
Section~\ref{rates} for all values of $(N \Gamma)_{\rm eff}(E)$ and
$\Gamma_N(E)$ considered below. Since the induced fission rates are
much smaller than the dipole absorption rate, we neglect fission in
this first approximation.  Furthermore, we approximate $(N
\Gamma)_{\rm eff}(E)$ by $N \Gamma_{\rm dip}$ for all values of $E$.
Beyond the critical energy $E_N$, neutron evaporation dominates
strongly, and it is practically impossible to excite nuclei to
energies larger than $E_N$. Since $E_N$ is smaller than the energy
$E_{\rm max}$ defined by the maximum of the level density, stimulated
photon emission is neglected.

Disregarding neutron evaporation we first determine the time and the
number of photons needed to reach the energy $E_N$ in the target
nucleus. We approximate the master equation for the target nucleus by
the set of equations
\be
\dot{P}(0, k, t) = - N \Gamma_{\rm dip} P(0, k, t) + N \Gamma_{\rm dip}
P(0, k - 1, t) \ .
\label{6}
\ee
The initial condition is $P(0, k, 0) = \delta_{k 0}$. The solution
\be
P(0, k, t) = \frac{(N \Gamma_{\rm dip} t)^{k}}{k!} \exp \{ - N
\Gamma_{\rm dip} t \}
\label{7}
\ee
obeys $\sum_{k = 0}^\infty P(0, k, t) = 1$ for all times $t$.
Considering $P(0, k, t)$ for fixed $t$ as a function of $k$ and using
Stirling's formula, we find that $P(0, k, t)$ has a maximum at $k =
k_{\rm max} = N \Gamma_{\rm dip} t$ with width $\sqrt{k_{\rm
    max}}$. The critical energy $E_N$ with $k(E_N) = k_N = E_N / E_L$
is reached after absorption of $N_0 = k_N \pm \sqrt{k_N}$ photons. The
time needed for the process is $t_N = E_N / (E_L N \Gamma_{\rm dip})$.
For $A = 100$, $E_N = 435$ MeV, $E_L = 5$ MeV, $N \Gamma_{\rm dip} =
5$ MeV that gives $N_0 = 87$ photons and $t_N = 12 \times 10^{- 21}$
s. The corresponding figures for $A = 200$ are $E_N = 1080$ MeV, $E_L
= 5$ MeV, $N \Gamma_{\rm dip} = 5$ MeV, $N_0 = 216$ photons and $t_N =
3 \times 10^{- 20}$ s. The laser pulse has the required length $t_N$
in time if $\sigma \leq 50$ keV or $\sigma \leq 22$ keV,
respectively. The speed of the process increases and the number of
absorbed photons decreases as either the photon energy $E_L$ or the
dipole absorption width $N \Gamma_{\rm dip}$ or both are increased.

As the process described by Eq.~(\ref{6}) carries on, neutron decay
actually depletes $P(0, k, t)$ and feeds $P(1, k', t)$. Even though
for energies below $E_N$ the neutron decay width $\Gamma_N$ is much
smaller than $N \Gamma_{\rm dip}$, the time needed to reach excitation
energy $E_N$ in the target nucleus is sufficiently large that nearly
the entire occupation probability in the target nucleus is lost to
neutron decay on the way. In the daughter nucleus, photon absorption
is described by the same equation~(\ref{6}) save for the feeding term.
Neutron decay is dominated by slow neutrons and, therefore, implies a
loss of excitation energy of about $8$ MeV (the binding energy).
Therefore, feeding of $P(1, k, t)$ by neutron decay occurs at $k$
values that are about two units (assuming a photon energy of around 5
MeV) smaller than the ones for which the loss in $P(0, k, t)$ occurs.
It takes about two absorbed photons to make up for that energy
loss. The daughter nucleus decays in turn by neutron emission. The
process repeats itself in the second daughter nucleus and so on. As a
result, the maximum of the occupation probability moves to ever
proton-richer nuclei and to ever higher excitation energies,
eventually hovering near or at most one or two $k$ units below $E_N$.

An overestimate for the time needed to reach the nucleus with mass
number $(A - i)$ at energy $E_N$ is obtained by disregarding neutron
decay in the target nucleus at energies below $E_N$, i.e., by using
Eq.~(\ref{6}) up to $E = E_N$, and by assuming that at $E = E_N$
neutron decay takes over and populates the states $(1, k_N - 2)$ in
the nucleus $(A - 1)$ (this scenario is consistent with $E_L=5$~MeV).
These are dipole-excited twice till the next neutron decay sets in,
and so on. Since at $E = E_N$ we have $\Gamma_N(E_N) = N \Gamma_{\rm
  dip}(E_N)$, the part of the master equation describing the feeding of
the first daughter nucleus by neutron decay of the target nucleus
reads
\ba
\dot{P}(1, k_N - 2, t) &=& - N \Gamma_{\rm dip} P(1, k_N - 2, t) \nonumber \\
&+& N
\Gamma_{\rm dip} P(0, k_N, t) \ .
\label{8}
\ea
The loss term accounts for photon absorption in nucleus $(A - 1)$.
From here Eq.~(\ref{6}) with $i = 0$ replaced by $i = 1$ takes over
until the energy $E_N$ is reached in nucleus $(A - 1)$. That is the
case after absorption of two photons. The process continues,
alternating between Eq.~(\ref{6}) and the analogue of Eq.~(\ref{8})
for the second, third, \ldots, daughter nucleus. Except for the
counting of $k$ values, these combined equations all have the same
form as Eq.~(\ref{6}), and their solution, therefore, has the form of
the right-hand side of Eq.~(\ref{7}). Hence
\be
P(i, k_n, t) = \frac{(N \Gamma_{\rm dip} t)^{k_n + 3 i}}{(k_n + 3 i)!}
\exp \{ - N \Gamma_{\rm dip} t \} \ .
\label{9}
\ee
The function $P(i, k_n, t)$ has its maximum at $t = (k_n + 3 i) / (N
\Gamma_{\rm dip})$. In other words, while it takes $87$ ($216$)
photons to reach $E_N$ from the ground state of the target nucleus
with mass number $A = 100$ ($A = 200$, respectively), it takes only
$15$ additional absorbed photons to move $5$ mass units away from the
line of stability. The additional time needed is $15 / (N \Gamma_{\rm
  dip})$. These figures show that once the threshold energy $E_N$ for
significant neutron evaporation is reached, the process quickly
populates nuclei far from the valley of stability. The spread $\sigma$
of the laser pulse needed to reach the energy $E_N$ is given by
\be
\sigma_0 = N \Gamma_{\rm dip} E_L / E_N \ .
\label{10}
\ee
Here $E_N$ is independent of  $E_L$. With
$E_N = 435$ MeV and $E_N = 1080$ MeV for mass numbers $A = 100$ and $A
= 200$, respectively, Eq.~(\ref{10}) defines a relation between the
experimental parameters $\sigma$, $E_L$, and $N \Gamma_{\rm dip}$. The
formation of proton-rich nuclei sets in whenever $\sigma \leq
\sigma_0$. Clearly, these times are overestimates because neutron
decay is actually simultaneous with (and not subsequent to) photon
absorption. After several ten zs, fission puts an end to the process,
sooner for heavy nuclei than for medium-weight ones.

%%%%%%%%%%%%%%%%%%%%%%%%%%%%%%%%%%%%%%%%%%%%%%%
\section{Numerical Results \label{numres}}
%%%%%%%%%%%%%%%%%%%%%%%%%%%%%%%%%%%%%%%%%%%%%%%

We calculate the time-dependent occupation probabilities $P(i, k, t)$
for medium-weight ($A = 100$) and heavy ($A = 200$) target nuclei that
interact with a short pulse of coherent MeV photons. We solve
Eq.~(\ref{1}) numerically for several choices of photon energy $E_L$,
of the effective dipole width $N \Gamma_{\rm dip}$, and of the length
$(n + 1)$ of the decay chain.  Eq.~(\ref{1}) is written in matrix form
(including target and daughter index $i = 0, 1, \ldots, n$) as
$\dot{\mathcal{P}} = \mathcal{M} \mathcal{P}$ with $P(i, k, t) \to
\mathcal{P}(k + i (k_{\rm max} + 1),t)$ and $k_{\rm max}$ the maximum
number of excitation steps considered. The matrix $\mathcal{M}$ is
independent of time and has both block-diagonal parts (fixed index
$i$) describing dipole absorption, stimulated emission, neutron decay
and fission, as well as non-diagonal feeding terms for the daughter
nuclei [mass $(A - i)$] which are populated by neutron decay of their
predecessors [mass $(A - i + 1)$]. The formal solution of
Eq.~(\ref{1}) in vector form is $\mathcal{P}(t) = \exp \{ \mathcal{M}
t \} \mathcal{P}(t = 0)$. Because of the strongly varying rates and
the large level densities involved (for instance, $\rho(E_{\rm max})
\simeq 10^{104}$~MeV$^{-1}$ for $A = 200$), severe numerical problems
may arise when one attempts to use standard linear algebra routines
such as LAPACK \cite{Lap99} for the calculation of the matrix
exponential.

\begin{figure*}[ht]
\includegraphics[width=0.33\linewidth]{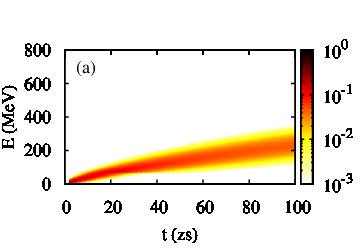}
\includegraphics[width=0.33\linewidth]{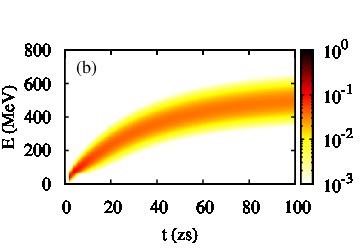}
\includegraphics[width=0.33\linewidth]{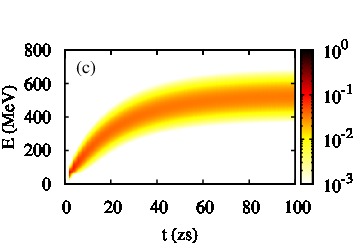}
\includegraphics[width=0.34\linewidth]{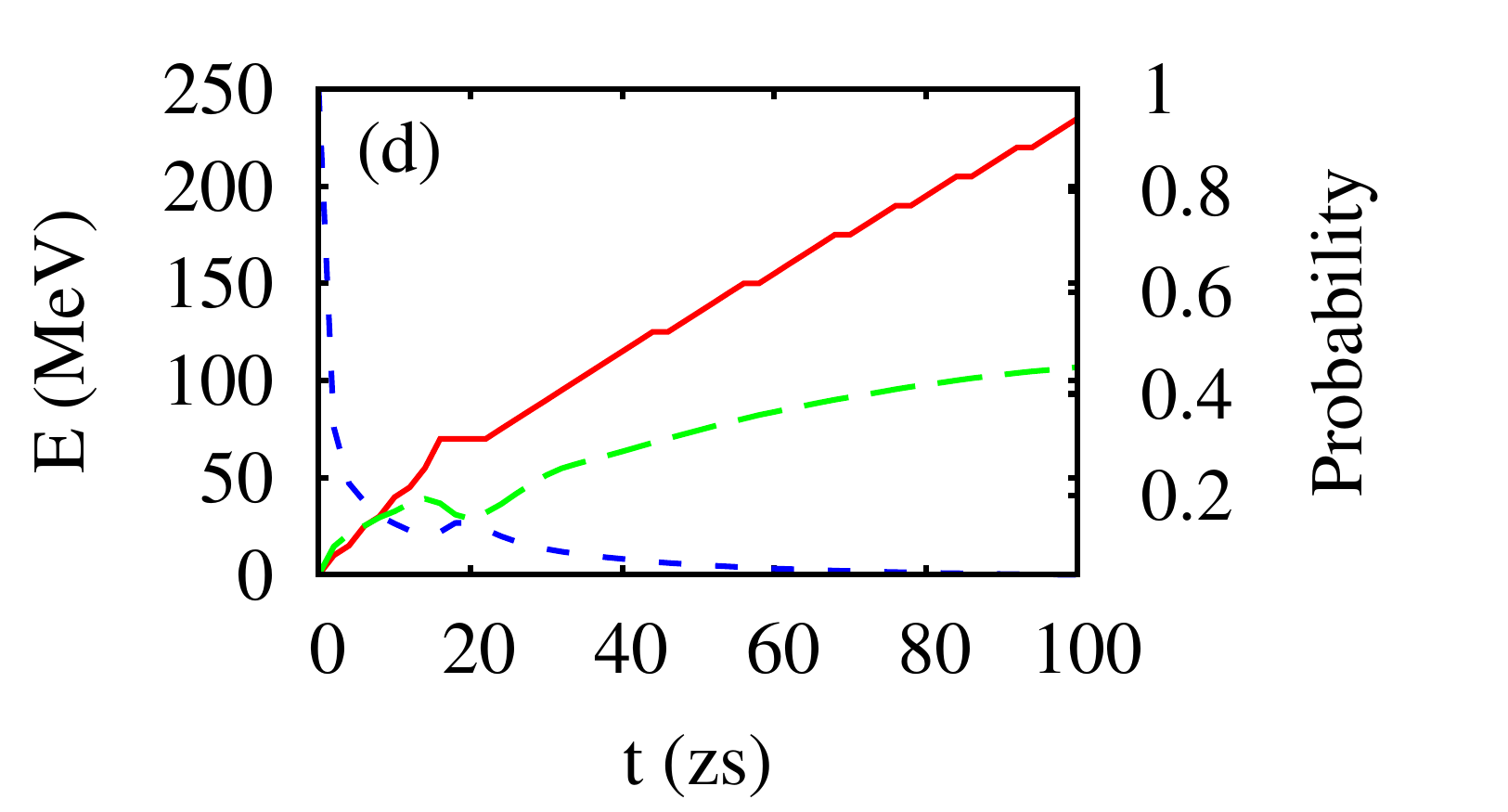}
\hspace{-0.33cm}
\includegraphics[width=0.34\linewidth]{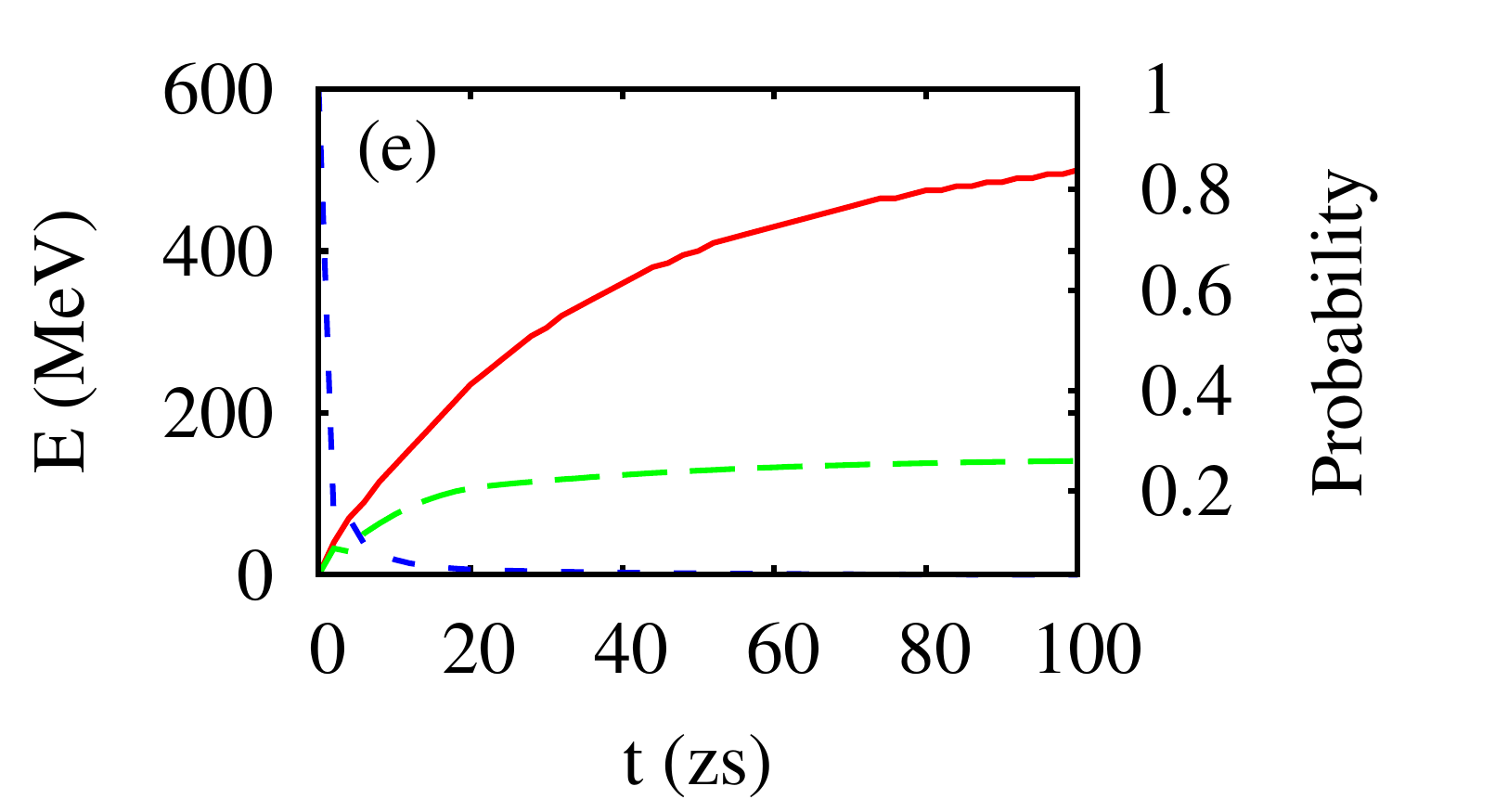}
\hspace{-0.33cm}
\includegraphics[width=0.34\linewidth]{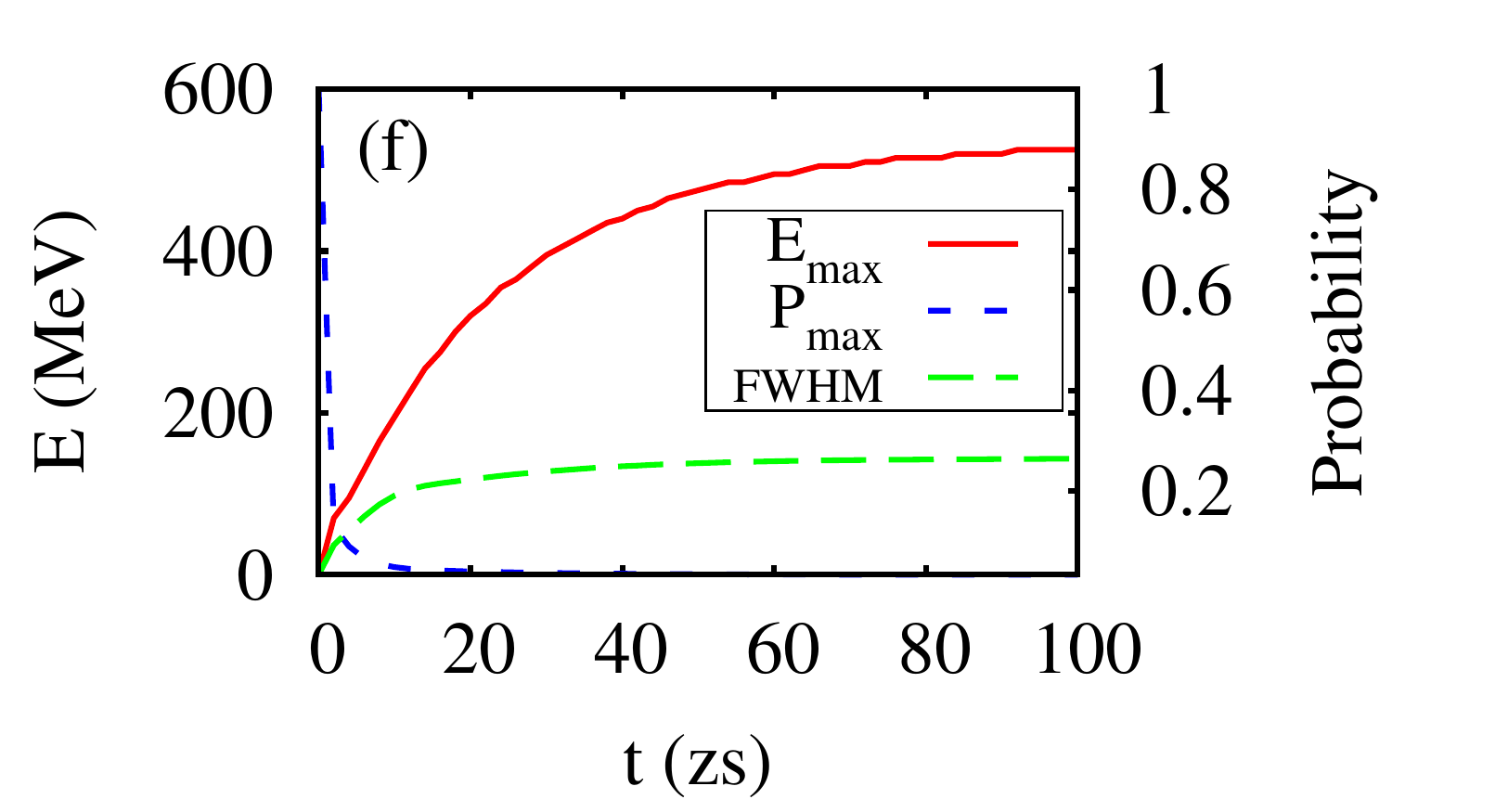}

\caption{(color online). Occupation probabilities $P(0, E, t)$ for
  $E_L = 5$~MeV, $A = 100$ and for dipole absorption and stimulated
  emission only. (a-c) Contour plots of the time-dependent occupation
  probability $P(0, E, t)$ as a function of excitation energy $E$ for
  (a) $N \Gamma_{\rm dip} = $1~MeV, (b) $N \Gamma_{\rm dip} = $5~MeV
  and (c) $N \Gamma_{\rm dip} =$ 8~MeV. (d-f) Corresponding peak
  height (dashed blue line) (scale on the right $y$-axis), position
  (solid red line), and FWHM (long-dashed green line) of the
  occupation probability (scales on the left $y$-axis) for (d) $N
  \Gamma_{\rm dip} =$ 1~MeV, (e) $N \Gamma_{\rm dip} =$ 5~MeV and (f)
  $N \Gamma_{\rm dip} =$ 8~MeV.}
\label{P5MeVA100}
\end{figure*}

Dedicated matrix exponential methods for solving systems of extremely 
stiff differential equations, i.e., equations for which the solving methods
 are numerically unstable, unless the step size 
is taken to be extremely small, have been developed in the last half-century 
\cite{Cody69,Hai96,Mol03} and applied, e.g., to nuclear fuel burnup 
calculations \cite{Pus10,Pus11,Stan12}. In this work, we employ a new fast 
general-purpose semi-analytical matrix exponential solver, which is based on 
a combination of backward-stable matrix algorithms \cite{Axel}.  First,
the system of differential equations is brought to triangular form using
Schur decomposition. Then, the triangular system is solved recursively using
a combination of Schur decomposition and back substitution algorithms. This
yields the solution in analytical form as a combination of polynomials and
exponential functions of time $t$. To verify 
numerical accuracy, we additionally carry out comparative calculations with 
other state-of-the-art solvers for the particularly demanding case of heavy nuclei with $A=200$.

Fig.~\ref{P5MeVA100} shows the target occupation probability $P(0, E,
t)=P(0,kE_L,t)$ in the absence of both neutron decay and fission for photon energy
$E_L = 5$~MeV and for $A = 100$ and $N \Gamma_{\rm dip} = $1, 5, 8
MeV. Depending on the effective dipole width, the saturation energy
$E_{\rm max} = 533$ MeV where stimulated emission limits photon
absorption is reached after 50 zs for $N\Gamma_{\rm dip}=$ 8 MeV
(after 100 zs for $N\Gamma_{\rm dip}=$ 5 MeV, respectively).  The
energy $E_N$ above which $\Gamma_N > (N \Gamma)_{\rm eff}$ is reached
for the case of $N\Gamma_{\rm dip}=$ 5 MeV much later than the time estimated in Sec.~\ref{estimates} of
$t_N = 12$ zs. A neck-like artifact can be observed at the switching
point $E=68$ MeV in the calculation method for the level densities
$\rho(i,k)$ and is best visible in Fig.~\ref{P5MeVA100}a.  Additional
information is provided by value and position of the maximum and by
the FWHM of the occupation probability shown in the lower part of the
figure for the three cases considered. In accord with our estimates in
Sec.~\ref{estimates} the FWHM is proportional to $\sqrt{t}$ and, after
the first few zs, the peak height has a $1 / \sqrt{t}$ dependence. The
linear dependence of the maximum on time holds only for the smallest
of the three values of $N\Gamma_{\rm dip}$ and does so only for small
times. In all other cases stimulated dipole emission slows down the
linear increase. The switching point at $E=68$~MeV is also visible
here in the shape of kinks in the three curves illustrating the peak
position, height and FWHM of the occupation probability in
Fig.~\ref{P5MeVA100}d.

\begin{figure*}[ht]
\includegraphics[width=0.33\linewidth]{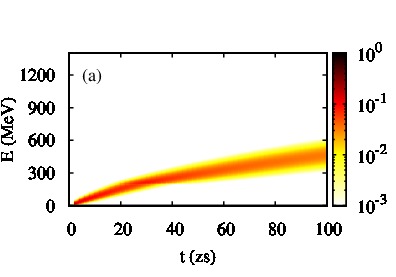}
\includegraphics[width=0.33\linewidth]{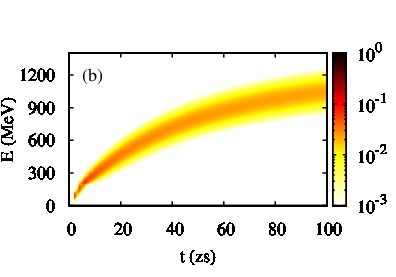}
\includegraphics[width=0.33\linewidth]{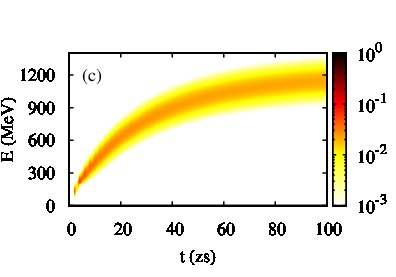}
\includegraphics[width=0.34\linewidth]{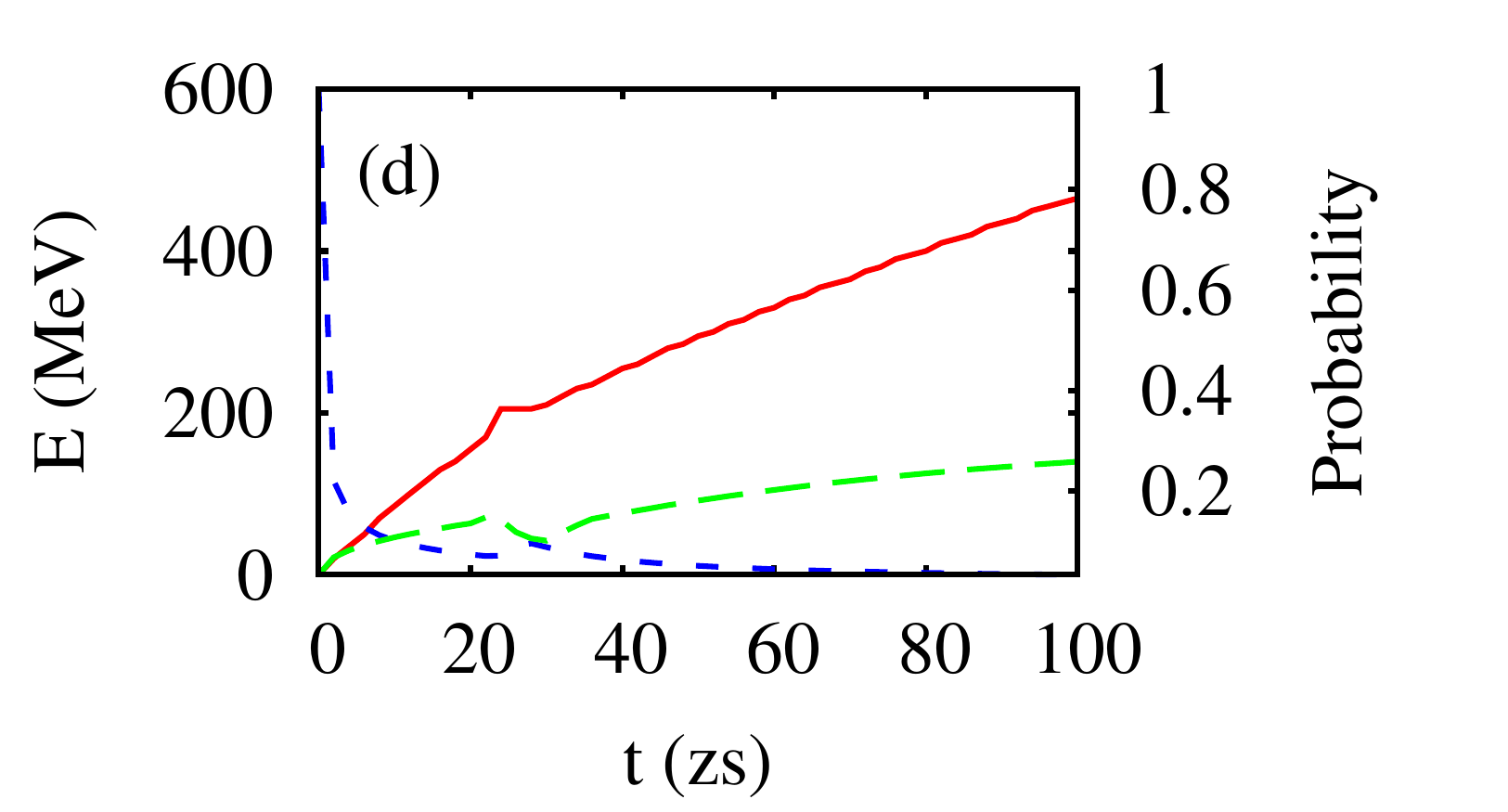}
\hspace{-0.33cm}
\includegraphics[width=0.34\linewidth]{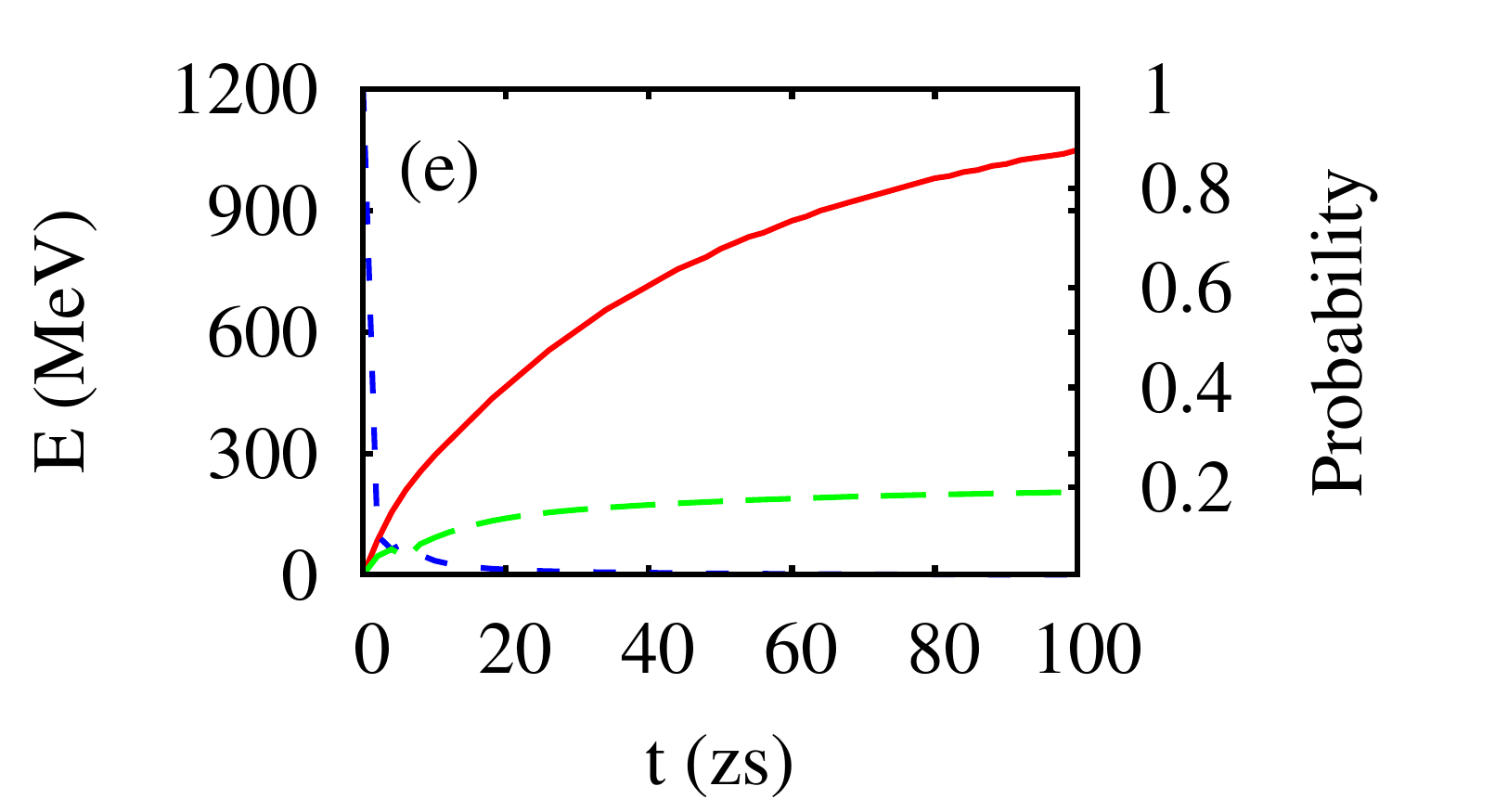}
\hspace{-0.33cm}
\includegraphics[width=0.34\linewidth]{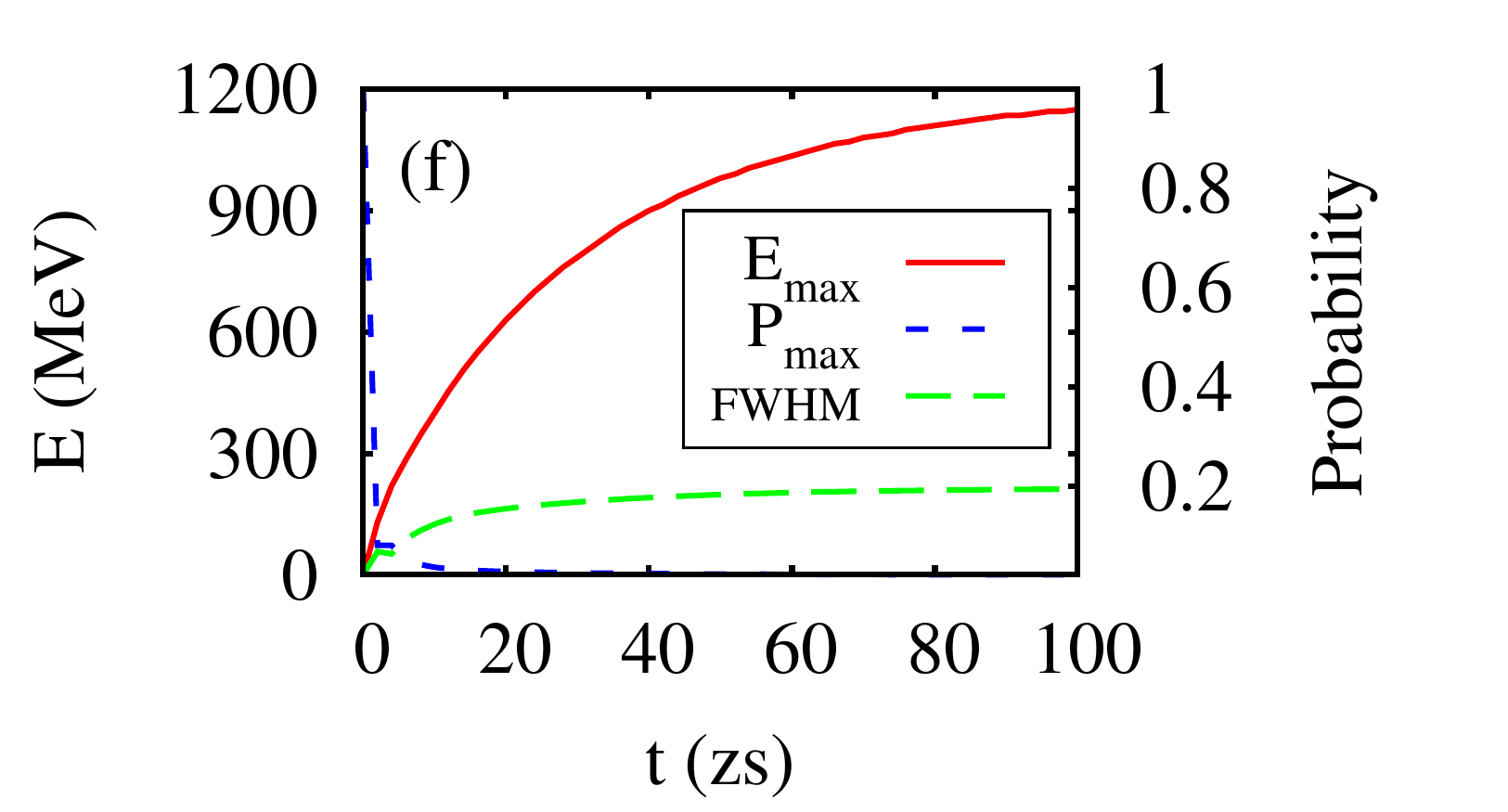}

\caption{(color online). Occupation probabilities $P(0, E, t)$ for
  $E_L = 5$~MeV, $A = 200$ and for dipole absorption and stimulated
  emission only. (a-c) Contour plots of the time-dependent occupation
  probability $P(0, E, t)$ as a function of excitation energy $E$ for
  (a) $N \Gamma_{\rm dip} = 1$~MeV, (b) $N \Gamma_{\rm dip} = 5$~MeV
  and (c) $N \Gamma_{\rm dip} = 8$~MeV. (d-f) Corresponding peak
  height (dashed blue line) (scale on the right $y$-axis), position
  (solid red line), and FWHM (long-dashed green line) (scales on the
  left $y$-axis) of the occupation probability and for (d) $N
  \Gamma_{\rm dip} = 1$~MeV, (e) $N \Gamma_{\rm dip} =5$~MeV and (f)
  $N \Gamma_{\rm dip} = 8$~MeV. }
\label{P5MeVA200}
\end{figure*}

Fig.~\ref{P5MeVA200} shows qualitatively similar results for a target
nucleus with $A = 200$. The time dependence of the peak position,
maximum peak value and FWHM again confirm our estimates in
Sec.~\ref{estimates}, except that the linear approximation for the
peak position is only valid for short times $t$ (except for the case
$N \Gamma_{\rm dip} =$ 1 MeV). Reaching the saturation energy $E_{\rm
  max} = 1200$ MeV requires a substantially larger number of absorbed
photons than for $A = 100$. The switching point between the
calculation methods for the level densities $\rho(i,k)$ at $E=200$ MeV
is also here visible, especially in Figs.~\ref{P5MeVA200}a and
\ref{P5MeVA200}d.

Laser excitation of heavy nuclei ($A = 200$) poses a numerically
challenging problem, due to increased stiffness of the equation
system. We have used the parameter set $A = 200$, $E_L = 5$ MeV, $N
\Gamma_{\rm dip} = 5$ MeV, as testing ground for five state-of-the-art
equation solvers that employ matrix exponential methods: (1) the 
semi-analytical matrix exponential method \cite{Axel}, (2) the 
Chebyshev Rational Approximation Method (CRAM)  with
partial fraction coefficients for approximation order
14 \cite{Pus11,Pus12}, (3) CRAM with partial fraction coefficients for
approximation order 16 \cite{Pus11,Pus12},  (4) eigenvector decomposition of
matrix $\mathcal{M}$ and  (5) a modern scaling and squaring Taylor expansion algorithm
\cite{scaling}.  The five solvers were used to
reproduce Fig.~\ref{P5MeVA200}b. The comparison shows that differently
calculated $P(0,E,t)$ values agree within an accuracy of $10^{-3}$.
All methods produce some numerical artifact in the form of fringes on
the upper side of  $P(0, k, t)$ on an accuracy level of
$10^{-4}-10^{-5}$. We do not expect that such a level of accuracy can
be attained in experiments in the foreseeable future and consider these
fringes  irrelevant. The semi-analytical algorithm used in the
present work~\cite{Axel} is the fastest one, yielding results in less than a
minute, while the slowest scaling and squaring Taylor expansion method required more than two weeks for the
same parameter set.

\begin{figure*}[ht]
\includegraphics[width=0.33\linewidth]{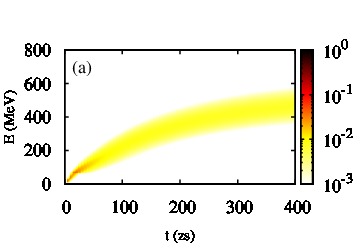}
\includegraphics[width=0.33\linewidth]{fig4b.jpg}
\includegraphics[width=0.33\linewidth]{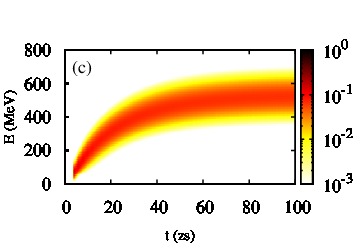}
\caption{(color online). Contour plots of the time-dependent
  occupation probability $P(0, E, t)$ as a function of excitation
  energy $E$ for (a) $E_L = 1$~MeV, (b) $E_L = 5$~MeV and (c) $E_L =
  10$~ MeV. The results are for $N \Gamma_{\rm dip} = 5$~MeV, $A =
  100$, and for dipole absorption and stimulated emission only.}
\label{P1-5-10MeVA100}
\end{figure*}

While Fig.~\ref{P5MeVA100} shows how the value of $N
  \Gamma_{\rm dip}$ affects the speed of nuclear excitation,
  Fig.~\ref{P1-5-10MeVA100} displays the influence of $E_L$. We take
  $E_L =$1, 5 and 10~MeV and use the same dipole absorption width $N
  \Gamma_{\rm dip} = 5$~MeV throughout. The dependence of the
  occupation probability $P(0, E, t)$ on photon energy $E_L$ is shown
  in Fig.~\ref{P1-5-10MeVA100} for the generic medium-weight target
  nucleus with $A = 100$ and in the absence of neutron decay.
Depending on photon energy, the saturation region is reached after 400, 100 or 50 zs,
respectively. A large photon energy speeds up the excitation process
and may partially counteract the effect of a small dipole absorption
width $N \Gamma_{\rm dip}$. For $E_L = 1$~MeV the excitation path is
energetically more narrow than in the other cases. In summary,
photon energy $E_L$ and dipole absorption width $N \Gamma_{\rm dip}$, 
(i.e., the number of coherent photons in the laser pulse)
jointly determine the time scale of the excitation process.
We add few technical remarks: Due to the smaller
excitation energy per step of the calculation, the case $E_L = 1$~MeV
is more sensitive to where we switch from the Bethe formula to our
method for the calculation of nuclear level densities. That point
shows up as a small neck at $\approx 68$~MeV excitation energy in all
three plots in Fig.~\ref{P1-5-10MeVA100}. In addition, the numerical effort for $E_L = 1$~MeV is much
larger than for higher values of $E_L$ as it requires both a large
matrix and, because of the slower excitation process, many more time
points. The difference in the energy spacing also affects the contour plots,
creating the false impression that the integral over occupation
probability at any one time is smaller for small $E_L$. Numerically, for any time $t$
the sum over all occupation probabilities $\sum_k P(0, k, t)$ equals unity with accuracy better 
than $10^{-6}$ for all three considered photon energies $E_L$.

\begin{figure*}[ht]
% target
\includegraphics[width=0.33\linewidth]{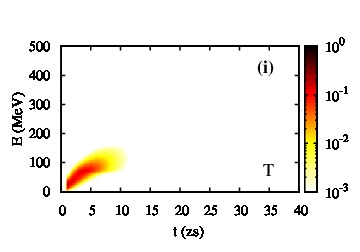}
\includegraphics[width=0.33\linewidth]{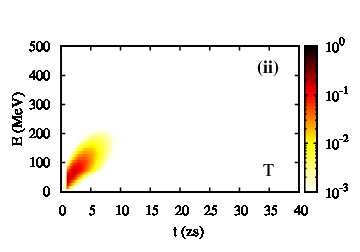}
\includegraphics[width=0.33\linewidth]{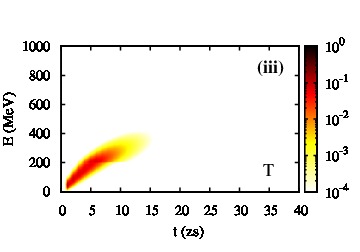}
% daughter 1
\includegraphics[width=0.33\linewidth]{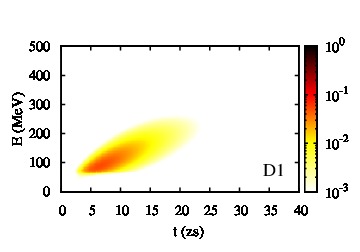}
\includegraphics[width=0.33\linewidth]{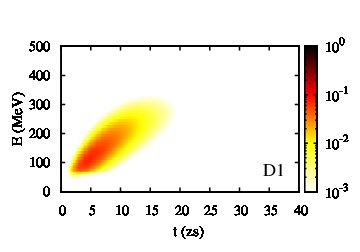}
\includegraphics[width=0.33\linewidth]{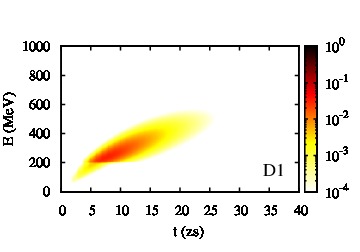}
% daughter 2
\includegraphics[width=0.33\linewidth]{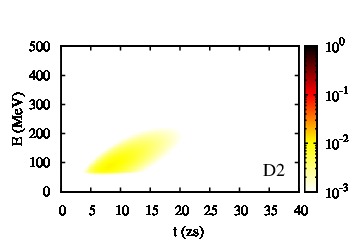}
\includegraphics[width=0.33\linewidth]{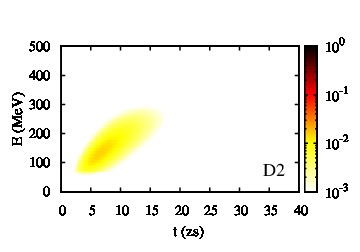}
\includegraphics[width=0.33\linewidth]{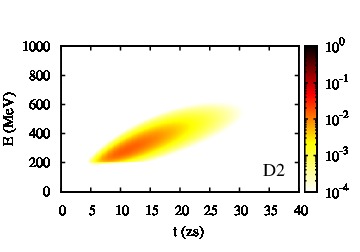}
% daughter 3
\includegraphics[width=0.33\linewidth]{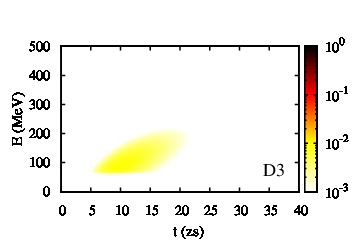}
\includegraphics[width=0.33\linewidth]{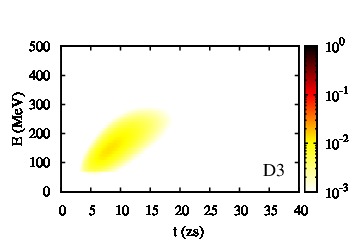}
\includegraphics[width=0.33\linewidth]{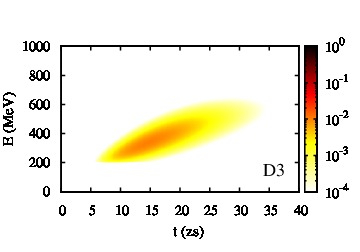}
% daughter 4
\includegraphics[width=0.33\linewidth]{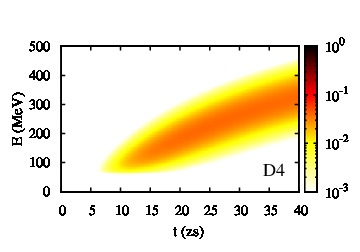}
\includegraphics[width=0.33\linewidth]{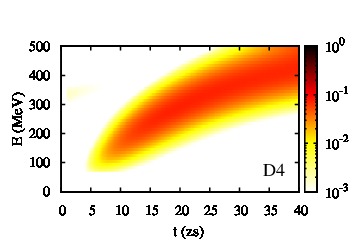}
\includegraphics[width=0.33\linewidth]{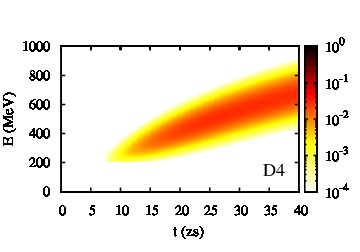}

\caption{(color online). Contour plots of the time-dependent
  occupation probabilities $P(i, E, t)$ with (from top to bottom)
  target nucleus ($i = 0$, label $T$) and four daughter nuclei ($i = 1$
  to $4$, labels $D1 -D4$) as functions of excitation energy $E$ for
  $N\Gamma_{\rm dip} =$ 5~MeV. Left column: Target nucleus $A = 100$,
  photon energy $E_L$ = 5~MeV; middle column: Target nucleus $A =
  100$, photon energy $E_L$ = 10~MeV; right column: Target nucleus $A
  = 200$, photon energy $E_L$ = 5~MeV. Please note the different color
  coding span for the case with target nucleus $A = 200$, which we
  chose for purpose of comparison with the next figure.}
\label{P4D}
\end{figure*}

We now take account of neutron decay of the target and of three
consecutive daughter nuclei, still neglecting fission. We solve the
master equation for a chain of five nuclei with mass numbers ranging
from $A$ to $A - 4$. We disregard neutron emission by the last nucleus
with mass number $A - 4$ which serves as a dump for the overall
probability flow. The dimensions of the matrices $\mathcal{M}$ are
five times larger than for the parent nucleus only. In particular, for
photon energy $E_L = 5$~MeV the matrix dimension is 500 (1000) for an
$A = 100$ (an $A = 200$) target nucleus, respectively. Fig.~\ref{P4D}
shows contour plots of the occupation probabilities $P(i, E, t)$ for
$i = 0$ (target) and $i = 1, 2, 3, 4$ (daughters) and for three
parameter sets as indicated. In all cases, the final nucleus in the
chain undergoes only dipole excitation with energies eventually
reaching the saturation energy. Compared to pure dipole absorption,
neutron emission is seen to broaden the distribution,
cf. Figs.~\ref{P5MeVA100}b and \ref{P5MeVA200}b. The occupation
probabilities $P(i, E, t)$ with $i \leq 3$ show that neutron emission
comes into play early, slowing down the excitation process even at
energies below $E_N$, i.e., below $\approx 435$~MeV for $A = 100$ and
1080 MeV for $A = 200$, respectively. As soon as the neutron emission
rates reach approximately $10^{20}$~s$^{-1}$, sufficiently many daughter
nuclei are produced within the tens of zs time span of the laser pulse
to strongly deplete the occupation probability of the target nucleus.
A comparison of the probability distributions of $A = 100$ parent
nuclei for photon energies $E_L$ of 5 and 10 MeV energy [columns (i)
and (ii) in Fig.~\ref{P4D}] shows that higher excitation energies are
reached (and correspondingly stronger neutron decay sets in) as the
photon energy is increased. For $t = 10$~zs and for $E_L = 10$~MeV
most of the nuclei have disintegrated to the dump $(i = 4)$, while for
$E_L = 5$~keV the target nucleus and the first daughter $i = 0, 1, 2$
are still predominantly occupied at $t<10$~zs. Comparing columns (i)
and (iii) of Fig.~\ref{P4D} we note that for identical times $t$,
higher excitation energies are reached in the heavier target.  That is
possible because of the much higher value of $E_N = 1080$~MeV for $A =
200$. As is the case for $A = 100$, neutron emission plays an
important role long before the excitation energy $E_N$ is
reached. After $10$ zs, the occupation probability of the target
nucleus practically vanishes, while the occupation probabilities for
nuclei with $i = 2$, 3 and even $i = 4$ are significant. 

\begin{figure*}[ht]
% parents
\includegraphics[width=0.33\linewidth]{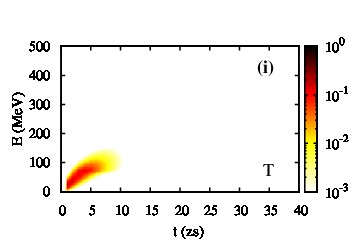}
\includegraphics[width=0.33\linewidth]{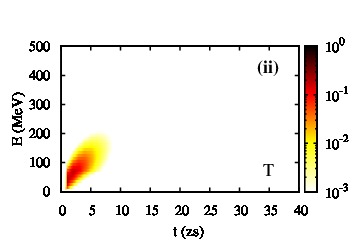}
\includegraphics[width=0.33\linewidth]{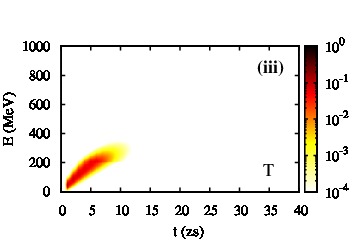}
% daughter 1
\includegraphics[width=0.33\linewidth]{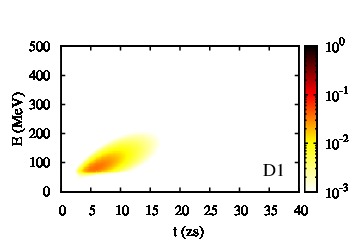}
\includegraphics[width=0.33\linewidth]{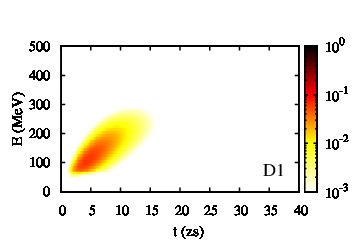}
\includegraphics[width=0.33\linewidth]{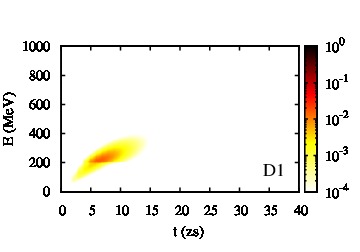}
% daughter 
\includegraphics[width=0.33\linewidth]{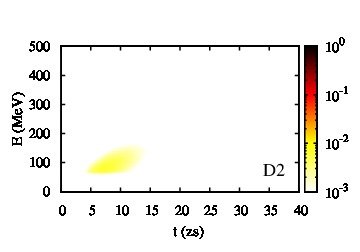}
\includegraphics[width=0.33\linewidth]{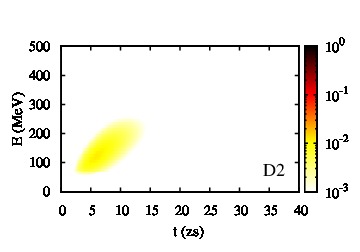}
\includegraphics[width=0.33\linewidth]{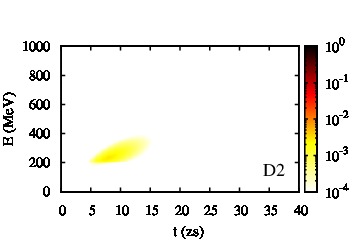}
% daughter 3
\includegraphics[width=0.33\linewidth]{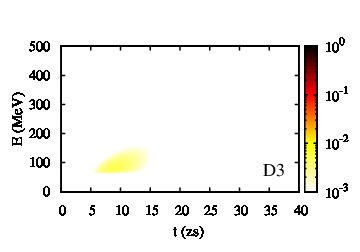}
\includegraphics[width=0.33\linewidth]{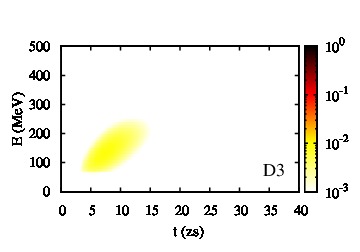}
\includegraphics[width=0.33\linewidth]{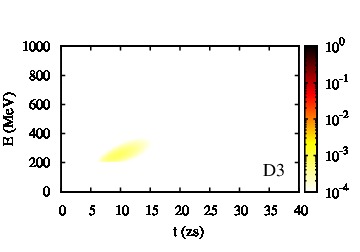}
% daughter 4
\includegraphics[width=0.33\linewidth]{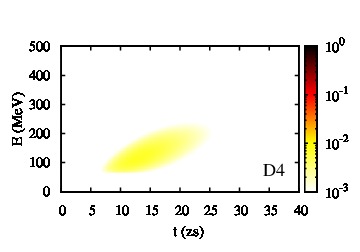}
\includegraphics[width=0.33\linewidth]{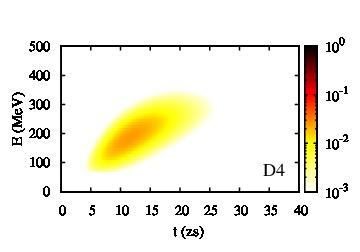}
\includegraphics[width=0.33\linewidth]{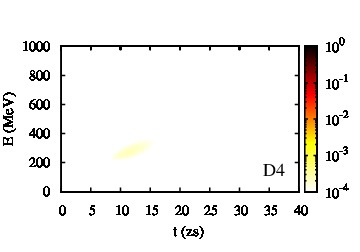}

\caption{(color online). Same as Fig.~\ref{P4D} but including induced
  fission.  }
\label{FissP4D}
\end{figure*}

\begin{figure}[ht]
\includegraphics[width=\linewidth]{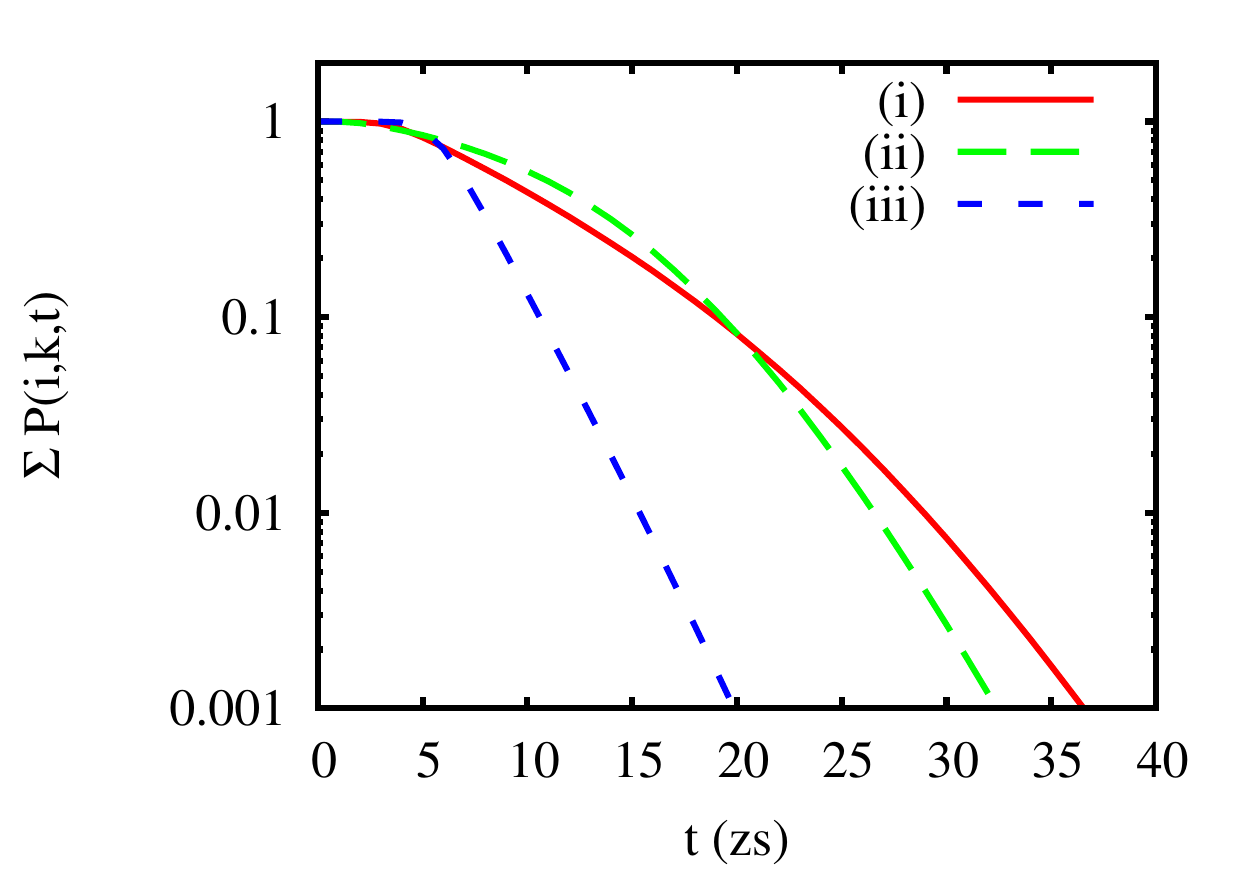}
\caption{(color online) Loss of total occupation
    probability $\sum_{i, k} P(i, k, t)$ due to fission for the three
    parameter sets used in Figs.~\ref{P4D} and \ref{FissP4D}:
    $N\Gamma_{\rm dip} =$ 5~MeV in all cases, (i) target nucleus with
    $A = 100$, photon energy $E_L$ = 5~MeV (solid red line), (ii)
    target nucleus with $A = 100$, photon energy $E_L$ = 10~MeV
    (long-dashed green line), and (iii) target nucleus with $A = 200$,
    photon energy $E_L$ = 5~MeV (short-dashed blue line).}
\label{Fissloss}
\end{figure}

 Fission is taken into account by adding the fission width as a
 diagonal loss term in Eq.~(\ref{1}). We do not follow the fate of the
 fission products as their masses are much smaller than those of the
 target nucleus and of the first few daughter nuclei. For a one-to-one
 comparison, results are shown in Fig.~\ref{FissP4D} for the same set
 of parameters as used in Fig.~\ref{P4D}. As expected from the decay
 rates shown in Sec.~\ref{rates}, Fig.~\ref{FissP4D} shows that
 neutron emission is much faster than fission. For $A = 100$, fission
 is so much slower than neutron decay that the comparison of the
 results for the target (T) and the first three daughter (D1-D3)
 nuclei shows little difference between the cases without and with
 fission. Only for the last nucleus in the chain, where neutron decay
 is switched off, we do reach the time scale for which the fission
 rate produces significant loss. Since heavy nuclei ($A = 200$) have a
 larger fission width, some loss of probability can be observed
 already for the first daughter nucleus and for the daughters with $i =
 3$ and $i = 4$ the occupation probabilities almost vanish. The
 effective loss of total occupation probability $\sum_{i, k} P(i, k,
 t) = 0$ is displayed in Fig.~\ref{Fissloss} for all three cases (i),
 (ii) and (iii).

%%%%%%%%%%%%%%%%%%%%%%%%%%%%%%%%%%%%%%%%%%%%%%
\section{Discussion \label{disc}}
%%%%%%%%%%%%%%%%%%%%%%%%%%%%%%%%%%%%%%%%%%%%%%

The results of Section~\ref{numres} are obtained under neglect of the
direct emission of nucleons by photoabsorption into the continuum. As
shown in Ref.~\cite{Pal14}, such direct emission plays only a minor
role for nuclei around $A = 100$ but is competitive with neutron decay
for heavy nuclei. The effective charges of neutrons and protons being
nearly equal in magnitude, such direct photoionization is expected to
produce neutrons and protons in about equal numbers. As a consequence,
photoabsorption populates highly excited states not only in the chain
of nuclei reached by neutron emission, but also in all nuclei with
mass numbers $(A - i)$ that lie between the valley of stability and
nuclei in that chain. We have not attempted to calculate that process
in detail.

Fission ultimately terminates all the processes considered in this
paper. Characteristic time scales are given in Fig.~\ref{Fissloss}.
The study of nuclei at high excitation energies and far off the line
of stability is possible only when the laser pulse terminates before
that characteristic time, i.e., if $\sigma \geq \Gamma_f$. It was
mentioned above that we expect $\Gamma_f$ to increase with increasing
distance of the fissioning nuclei from the valley of stability. A
reliable estimate would require a precise calculation of the height of
the fission barrier $E_f$ versus that distance. Termination by fission
of the processes considered in this paper would allow to measure
$\Gamma_f$ and, thus, to check such calculations.

It was pointed out by A. Richter~\cite{Ric14}  that at excitation 
energies of several $100$ MeV, the compound
nucleus might undergo transitions in which excited states of the
nucleon are populated or in which subthreshold pion production
occurs. How likely are such processes (which we have disregarded in
this paper)? In the quasiadiabatic regime, the compound nucleus is near
equilibrium at all times, and the answer follows from a statistical
argument. The large number of equilibrium configurations at energy $E$
must be compared with the very much smaller number of configurations
where (almost) all the energy resides in a single mode (that of the
excited nucleon or that of the pion). The situation is comparable to
the emission of a fast neutron that carries almost all the excitation
energy of the compound nucleus and leaves the residual nucleus in a
state of low excitation energy. In comparison with slow-neutron
emission, this process is suppressed by up to twenty orders of
magnitude. We expect a large suppression factor also for the processes
under discussion. However, each of these processes is easily
distinguishable experimentally from any of the processes considered in
our paper. In spite of their enormous scarcity, formation of any of
the $\Delta$ resonances or subthreshold pion production might,
therefore, still be observable.

 In conclusion, in the quasiadiabatic regime reactions
  induced by coherent laser light will cause the absorption of up to
  several $100$ photons. Such reactions offer the unique chance to
  explore the level density of the compound nucleus far above the
  yrast line, hitherto an unknown territory. The model we have used is
  plausible but not yet established. It is for experimental data to tell
whether our present understanding is correct. The reaction will
  produce nuclei far from the valley of stability. Their decay
  processes will yield spectroscopic information not accessible so far.

%%%%%%%%%%%%%%%%%%%%%%%%%%%%%%%%%%%%%%%%%%%%%%%%%%%%%%%%%%%%%%%%%%%%%%%%%%
\appendix
\section{\label{A}}
%%%%%%%%%%%%%%%%%%%%%%%%%%%%%%%%%%%%%%%%%%%%%%%%%%%%%%%%%%%%%%%%%%%%%%%%%%

We calculate the distribution of spin values in the target nucleus
after absorption of $N_0$ photons. We take the direction of
propagation of the photons as $z$--axis. In an unpolarized laser beam
each dipole photon carries the $z$-component angular momentum $\pm 1$. Absorption of a single photon populates nuclear states with $J_z =
\pm 1$. These correspond to total spin $J = 1$. Absorption of two
photons populates nuclear states with $J_z = \pm 2$ (once each) and
$J_z = 0$ (twice), corresponding to $J = 2$ (once) and $J = 0$
(once). Absorption of $N_0$ photons populates states with $J_z = \pm
N_0, \pm N_0 \mp 2, \pm N_0 \mp 4, \ldots$. Inspection shows that the
multiplicity of states with given $|J_z|$ is given by ${N_0 \choose
  k}$ where $2 k = |J_z| + N_0$ and $k = N_0/2, N_0/2 + 1, \ldots,
N_0$ for $N_0$ even and $k = (N_0 + 1)/2, (N_0 + 3)/2, \ldots, N_0$
for $N_0$ odd. We note that for $N_0$ even (odd) only even (odd)
values of $J_z$ and, thus, of total spin $J$ occur. The number $Z(J)$
of such spin values is given by ${N_0 \choose k} - {N_0 \choose k +
  1}$ with $2 k = J + N_0$. For $N_0 \gg 1$ we use Stirling's formula
and approximate the difference by the negative derivative with respect
to $k$. Thus,
\begin{widetext}
\be
Z(J) = - \frac{\rm d}{{\rm d} k} {N_0 \choose k} \bigg|_{k = (N_0 +
J)/2} \approx \ln \frac{N_0 - J}{N_0 + J} \exp \{ N_0 \ln N_0 - (1/2)
(N_0 + J) \ln (N_0 + J)/2 - (1/2)
(N_0 - J) \ln(N_0 - J)/2 \} \ . 
\label{A1}
\ee
\end{widetext}
By definition, $Z(J) \geq 0$. We note that $Z(J) = 0$ at $J = 0$.
Inspection shows that $Z(J)$ has a single maximum and drops off to
relatively small values for $J \approx N_0$. The maximum $J_0$ of
$Z(J)$ is located at $J_0 = \sqrt{N_0}$. We have used $J_0 \ll N_0$.
For $N_0 \gg 1$ that condition is met by the solution~(\ref{A1}).

%%%%%%%%%%%%%%%%%%%%%%%%%%%%%%%%%%%%%%%%%%%%%%%%%%%%%%%%%%%%%%%%%%%%%%

\end{document}